\documentclass{emulateapj}  

\begin{document}

\title{THE X$^-$ SOLUTION TO THE $^6$Li AND $^7$Li BIG BANG
NUCLEOSYNTHESIS PROBLEMS}

\author{
 MOTOHIKO KUSAKABE\altaffilmark{1,2},
 TOSHITAKA KAJINO\altaffilmark{1,3},
 RICHARD N. BOYD\altaffilmark{4},
 TAKASHI YOSHIDA}

\affil{Division of Theoretical Astronomy, National Astronomical
Observatory of Japan, Mitaka, Tokyo 181-8588, Japan\\
{\tt kusakabe@th.nao.ac.jp}}

\altaffiltext{1}{Department of Astronomy, Graduate School of Science,
University of Tokyo,  Hongo, Bunkyo-ku, Tokyo 113-0033, Japan}
\altaffiltext{2}{Research Fellow of the Japan Society for the Promotion
of Science}
\altaffiltext{3}{Department of Astronomical Science, The Graduate
University for Advanced Studies, Mitaka, Tokyo 181-8588, Japan}
\altaffiltext{4}{Present Address: Lawrence Livermore National
Laboratory, 7000 East Avenue, Livermore, CA 94550, USA}

\and
\author{GRANT J. MATHEWS}

\affil{Department of Physics and Center for Astrophysics, University of
Notre Dame, Notre Dame, IN 46556, USA}

\begin{abstract}
The $^6$Li abundance observed in metal poor halo stars appears to exhibit a
 plateau as a function of metallicity similar to that for $^7$Li,  suggesting a big bang  origin.
 However,  the inferred primordial abundance of  $^6$Li is $\sim$1000 times larger
 than that predicted by  standard big bang nucleosynthesis for the baryon-to-photon ratio  inferred
 from the WMAP data. Also, the inferred $^7$Li primordial abundance is 3 times smaller than the big bang prediction.    We here describe in detail a possible simultaneous solution to both the problems of
 underproduction of $^6$Li and overproduction of $^7$Li in big bang
 nucleosynthesis.  This solution involves a hypothetical massive,
 negatively-charged leptonic particle that would bind to the light
 nuclei produced in big bang nucleosynthesis, but would decay long
 before it could be detected.   We consider only the $X$-nuclear reactions and assume that the effect of decay products is negligible, as would be the case if lifetime were large or the mass difference between the charged particle and its daughter were small. An interesting  feature of this paradigm is that,
 because the particle remains bound to the existing nuclei after the cessation of the usual big bang nuclear reactions, a second longer epoch
 of nucleosynthesis can occur among $X$-nuclei which have reduced Coulomb
 barriers.  The
 existence of the $X^-$ particles thus extends big bang nucleosynthesis
 from the first few minutes to roughly five hours.  
 We confirm that reactions in which the hypothetical particle is
 transferred can occur that greatly enhance the production of $^6$Li while depleting $^7$Li.  
 We also identify a new reaction that destroys large amounts of $^7$Be, and hence reduces the 
 ultimate $^7$Li abundance.  Thus,
 big-bang  nucleosynthesis in the presence of 
 these hypothetical particles, together with or without an
 event of stellar processing, can simultaneously solve the two
 Li abundance problems.
\end{abstract}


\section{INTRODUCTION}\label{sec1}

A long-standing effort in astrophysics has involved searches for
signatures of unstable particles that might have existed in the early
Universe, but have long since become extinct.  The precision that exists
in the abundances of the nuclides produced in  big bang
nucleosynthesis (BBN) suggests that primordial nucleosynthesis  is a good place to look for
such signatures.  In that context, it is of considerable interest that
recent observations~\citep[e.g.][]{asp06} of  
metal poor halo stars (MPHSs) indicate that the primordial abundances of 
both $^6$Li and $^7$Li  may  not be in agreement with the predictions of standard BBN.
 Specifically, the $^6$Li 
appears to  have an abundance plateau similar to that for $^7$Li in very low metallicity
stars.  This suggests a primordial abundance.  However the abundance value is
  roughly a factor of 1000 larger than predicted in standard BBN.  A similar, though
much less severe, problem exists for $^7$Li~\citep{rya00,mel04,asp06}; the BBN value that would be
consistent with the baryon-to-photon ratio inferred from analysis of the
Wilkinson Microwave Anisotropy Probe
(WMAP)-power spectrum suggests an abundance that
is roughly a factor of three higher than the 
observed value.

In response to this intriguing  $^6$Li result, a
 number of solutions have been suggested.  Some relate  
 the Li anomalies to the possible existence of unstable heavy
particles in the early
 Universe~\citep{dim88a,dim88b,dim89,jed00,cyb03,jed04,kaw05,kus06} which lead to $^6$Li production via non-thermal reactions induced 
by the particle decay.  Some have suggested a possible epoch of enhanced cosmic ray nucleosynthesis
via the $\alpha + \alpha$ reactions~\citep{rol06,pra06,kus07a}.   

    Of interest to the present work, however,  is the suggestion in several 
recent works~\citep{cyb06,kap06,pos07,koh07,ham07,kus07b,jedamzik08a,jedamzik08b} that heavy negatively charged
unstable particles could modify BBN and lead to $^6$Li production.  However, this mechanism
would operate in a  rather different way from the previous suggestions.  In
this paradigm, the heavy particles (here denoted as $X^-$) would bind to
the nuclei synthesized during BBN to produce exotic nuclei (hereafter denoted
as $X$-nuclei).  These massive $X^-$ particles would be bound in orbits that
would have radii comparable to those of the nuclei to which they were
attached.  Their presence, however,  would reduce the Coulomb barrier seen by incident charged
particles, thereby enhancing the thermonuclear reaction rates and allowing a longer duration of
BBN in an extended epoch of nucleosynthesis.  In the present work we
expand upon our previous study~\citep{kus07b} and describe details of a revised
analysis of $X^-$ particle effects on BBN.  The focus of the present work is to better quantify the effects of $X^-$ particles on BBN.  It is possible, however, that both $X^-$ effects and the $X$-particle decay contribute to the abundances.  In a future work we will consider this, but for now we only address the effects of $X^-$ particles and show that this alone can explain the observed lithium abundances.

In standard BBN,  $^6$Li  production is suppressed.  It is synthesized  
primarily via the $^4$He($d$,$\gamma$)$^6$Li reaction, which has a very small cross section.  However,~\citet{pos07}  suggested that a large enhancement of the $^6$Li abundance 
could result from an $X^-$ bound to $^4$He (denoted $^4$He$_X$).  This 
allows for the $X^-$ transfer reaction of $^4$He$_X$($d$,$X^-$)$^6$Li, which can enhance the  production of $^6$Li by as much as seven orders of
magnitude.  Subsequently,~\citet{ham07} carried out a theoretical
calculation of the cross section for this $X^-$ transfer reaction in a
quantum three-body model.  Their value was about an order of magnitude
smaller than that of~\citet{pos07}.  This difference can be traced  to their exact treatment of the
quantum tunneling in the fusion process and  the use of a better nuclear
potential.  

In other work,~\citet{cyb06} identified the $X^-$ as the
supersymmetric counterpart of a tau, i.e., a stau., thereby providing some plausibility  for
the existence of such particles.  They also considered the
$X^-$ transfer reactions for $^6$Li, $^7$Li, and $^7$Be production as
Pospelov suggested.  \citet{kap06} observed that the decay of an
 $X^-$ when bound to $^4$He  occasionally knocks out a
proton or neutron to produce $^3$He or $^3$H, thereby enhancing their abundances
in BBN.  Those nuclides could then interact with other primordial
$^4$He nuclei at higher energies than those normally associated with the BBN
production of  $^6$Li.  \citet{koh07} and~\citet{kus07b} studied the recombination of
nuclei with $X^-$ particles, and suggested the possibility that BBN with charged
massive particles also destroys $^7$Li and so solves the $^7$Li problem
as well.  In addition, a resonant reaction
$^7$Be$_X$($p$,$\gamma$)$^8$B$_X$ has recently been proposed (Bird et
al. 2007) that further destroys $^7$Be$_X$ through an atomic excited state of
$^8$B$_X$.

In the present work we have estimated new reaction rates for the $X^-$
transfer reactions suggested by~\citet{pos07} as well as others that could occur.  We have reanalyzed the $X^-$
transfer reaction rates estimated in~\citet{cyb06} and  found them to be negligible
based upon the reaction dynamics discussed below.

We
have adopted a simple model to estimate the binding energy due to the
bound $X^-$ particles for each nucleus involved in BBN.  The reaction
rates are corrected for the modified nuclear charges and the effective
mass resulting from the binding of the $X^-$ particle.  The reaction rates, however,   are not
particularly sensitive to the mass of the $X^-$ particle.   Therefore, we have
simply assumed that they are  much heavier than the nucleon mass.  

In this paper we study the potential consequences of $X^-$ particles on BBN,
including recombination of the $X^-$ particles by the existing normal
nuclei.  Several new reaction processes have been included that can produce
or destroy $^6$Li and $^7$Li in the early universe.  We utilize 
a fully dynamical description which is solved numerically, and which includes all relevant details of kinetic and chemical equilibrium.  The purpose of this present work is to
better quantify whether there are  observable consequences of the resulting
reaction network.  The ultimate goal is to better 
explain  the observed overproduction (1000 times) and underproduction (3 times)
of  the abundances of $^6$Li and $^7$Li.

In Section~\ref{sec2} we outline some of the details of the
calculations we have performed to estimate the thermonuclear
reaction rates, and to describe the evolution and abundance  of
the different nuclei, $X^-$ particles, and $X$-nuclei during the BBN.  

In Section~\ref{sec3}, we show the results of BBN with the
reactions involving the nuclides that could be formed with the embedded
$X^-$ particles.     As was done in~\citet{jedamzik08a}, we identify the parameter space for solving both the the $^6$Li
and $^7$Li BBN abundance problems.  The present work, however,  differs from that of
\citet{jedamzik08a} in several important ways.   Besides differences in some of the relevant reaction rates, there are two major differences.  One is the inclusion of $X$-transfer reactions.  \citet{jedamzik08a} found that $X$-transfer reactions involving the charge $Z = 1$ $X$-nuclei could possibly change the light-element abundances.  That result, however,  relied on  reaction rates which were calculated within the framework of the Born approximation.  Recently, however, a detailed study of one of the relevant $X$-transfer reactions by~\citet{ham07} shows that a more realistic calculation gives a much smaller (factor of 10)  $X$-transfer reaction rate.  Based upon that study, we conclude that $X$-transfer reactions can be neglected in this work.    Another difference  is the assumption that hadronic and electromagnetic (EM)  decays are unimportant.  That is, we only consider the $X$-nuclear reactions and assume that the effect the of decay products is negligible.  This is possible if the decay lifetimes are long and/or  the mass difference between the $X^-$  and daughter particle is small. In supesymmetric  scenarios for example~\citep[e.g.][]{feng03a,feng03b}, nearly equal masses  would also imply a larger lifetime. Very large lifetimes are ruled out but lifetimes of order a year  are very interesting for structure formation \citep{Sigurdson04}. 
 
 In Section~\ref{sec4}, the effects  $X^-$ decay are discussed.  Our conclusions are summarized in Section~\ref{sec5}.

\section{MODEL}\label{sec2}

In order to perform the nucleosynthesis calculations, we have added the
relevant $X$-nuclei and their reactions to the BBN network code~\citep{kawano}.  The $X$-nucleus $^4$He$_X$ is particularly important for
the present work, although the other
$X$-nuclei are included and can significantly affect the results.  Both proton and neutron
captures involving the $X^-$ particles were included when energetically allowed, as were
''transfer reactions.''   We modified
most of the thermonuclear reaction rates on the $X$-nuclei from the original
rates (without $X$-nuclei).  The two dominant effects were  the lowered Coulomb
barriers resulting from the $X^-$ in the nucleus, and the modified reduced
mass.  However, as noted below, there are a number of reactions for which 
careful additional consideration is required in order to estimate reliable reaction rates.

\subsection{Properties of the $X^-$ Particle}

The $X^-$ particle is assumed to be leptonic, since there is motivation
from particle physics for the existence of such a particle~\citep{cyb06}.  
Identifying these particles as 
the supersymmetric partners of normal leptons implies that they would also be leptonic but of
 spin 0.  Such sleptonic  $X^-$ particles will be initially produced in pairs with
$X^+$ particles.  Ultimately, their annihilations in the early universe will freezeout.
The final required BBN abundance of the residual $X^+$-$X^-$ pairs will then constrain their annihilation cross section.  We show below that this cross section is consistent with a weakly interacting 
sleptonic particle.
We note, that the $X^+$ particles, though also present during BBN will have negligible interactions with 
ambient nuclei (compared to $X^-$ particles) due to their Coulomb
repulsion and low associated reaction rate.   Hence, they can be
neglected in the present analysis.  It is possible, however, that they could affect the final results to the production of electromagnetic and/or hadronic showers when they decay.  That complication will be addressed in a future paper.  The focus of the present work, however, is only the effects of $X^-$ particles on BBN.

Ultimately, the product of
the $X^-$ abundance and mass must be consistent with the  WMAP CMB power spectrum.
The lifetime of the $X^-$ also has a lower limit because it must  live
long enough to allow at least some fraction of the $X^-$ produced
initially in the big bang to exist through the epoch of BBN.  If the
$X^-$ particles decay into leptons and photons at some later stage,
they are expected to destroy some fraction of the nuclei to which they
had become bound during BBN.  However, that fraction is not expected to
be large~\citep{kap06,ros75} and can be neglected.

\subsection{Nuclear Binding Energies}

The reaction rates of the  $X$-nuclei are strongly affected by their binding energies.
In our calculations both  the binding energies
and the eigenstate wave functions of $X^-$ particles were computed by taking into account the modified Coulomb interaction with the nucleus.  For this purpose, we assumed that
the charge distribution of nuclides is Gaussian.  We then solved the
two-body Shr\"{o}dinger equation by a variational calculation~\citep[Gaussian
expansion method, i.e.][]{hiy03}, and obtained binding energies.  The
obtained values are listed in Table \ref{tab1}.  The adopted root
mean square charge radii are listed in the second column.  When
experimental data exist, we used charge radii determined from experiment.  Note that since the $X^-$ particles can bound only
electromagnetically to nuclei, their binding energies are typically
small ($\sim 0.1-1$~MeV), and are largest for heavy nuclei.  Hence, they
are not appreciably bound to nuclei until the  temperature becomes low enough.

\subsection{Reaction Rates}

The leading term in the expression for the thermonuclear reaction rates (TRR) $\langle
\sigma v \rangle$ for nonresonant reactions involving  nuclei with
embedded $X^-$ particles can be  written~\citep[e.g.][]{boy07} as
\begin{eqnarray}
\langle \sigma v \rangle_{\rm NR} &=& \left(\frac{2}{\mu}\right)^{1/2} \left(k_B T\right) \left(\frac{4}{9}\right)3^{-1/2} \nonumber \\
&&\times E_0^{-3/2} S(E_0) \tau^2 \exp(-\tau),
\label{eq1}
\end{eqnarray}
where $E_0 = 1.22 (z_1^2 Z_2^2 \mu T_6^2)^{1/3}$ keV  is the energy at the
peak of the Gamow window, $S(E_0)$ is the ''astrophysical
$S$-factor'' at $E_0$, $k_B$ is the Boltzmann constant, $T_6$ is the
temperature in units of $10^6$~K, and
\begin{equation}
\tau = \frac{3E_0}{k_B T}= 42.46 \left(\frac{z_1^2 Z_2^2 \mu}{T_6}\right)^{1/3}.
\label{eq2}
\end{equation}
The astrophysical $S$ factor contains the nuclear matrix element for the reaction.
For some of the rates we take the nuclear matrix element within  $S(E_0)$ to be the same as for the reactions of  the
corresponding normal nuclei~\citep[NACRE;][]{cau88,smi93}.  For others we can adopt a scaling relation as described below.  Hence, the dominant corrections for the TRR
in the above equation arises from the terms involving the reduced mass  $\mu$ (in atomic mass units),
and $z_1$ and $Z_2$ (the  proton numbers for the projectile
normal nucleus and the target $X$-nucleus).  We have assumed $Z_2$ to be
the net charge of the bare nucleus and any embedded $X^-$ particles,
i.e. $Z_2=z_2-n$ for single ($n=1$) $X^-$-bound
nuclei, where $z_2$ is the atomic number of the bare nucleus.  As noted above, we take the spin
of the $X^-$ particles to be zero.

At the epoch during  BBN at which the $X^-$-nuclei  undergo nucleosynthesis
 the neutron abundance is extremely small.  Nevertheless, for completeness we have
   included neutron radiative
capture reactions in the reaction network.  We adopted the known values
for the normal nuclei whenever possible, but those values are not always
available for the unstable nuclei of interest $3 \leq z_2 \leq 5$.  
For the Be$_X$, B$_X$($n$,$\gamma$) reactions we assumed a mean value obtained from the
$^6$Li($n$,$\gamma$)$^7$Li and $^7$Li($n$,$\gamma$)$^8$Li
 reactions~\citep[see, e.g.][and references therein]{hei98,bar80} of 40~$\mu$b at 25 keV.  

\subsubsection{$X^-$ Transfer Reactions to Produce $^6$Li, $^7$Li and
   $^7$Be}

Reactions in which an $X^-$ particle can be
transferred can be very important in circumventing some of the reactions
that would normally be inhibited~\citep{pos07}.  This is especially true of the
$^4$He$_X$($d$,$X^-$)$^6$Li  reaction.  The rate for this reaction could be 
 orders of magnitude larger than that of the $^4$He($d$,$\gamma$)$^6$Li
reaction.  This latter reaction is the main process by which $^6$Li is made in BBN 
in the absence of $X^-$
particle.  Normally, however, this reaction is suppressed because it is 
dominated by an electric quadrupole
transition.  

\citet{ham07} have recently carried out a new
theoretical calculation of the cross section for the $X^-$ transfer reaction
$^4$He$_X$($d$,$X^-$)$^6$Li in the context of a quantum three-body model.  Their value
was about an order of magnitude smaller than that of~\citet{pos07}.
This difference can be traced to 
an  exact treatment of the quantum tunneling in the fusion
process  and the use of a better nuclear potential.  We have  therefore adopted
the rate of \citet{ham07}
 as the most reliable estimate for this rate and assume a factor
of 3 uncertainty.

\citet{cyb06} estimated astrophysical $S$-factors for various transfer  reactions including the
$^4$He$_X$($t$,$X^-$)$^7$Li,    $^4$He$_X$($^3$He,$X^-$)$^7$Be,
$^6$Li$_X$($p$,$X^-$)$^7$Be reactions by applying a scaling
relation~\citep{pos07}, 
\begin{equation}
S_X/S_\gamma \propto p_fa_0/(\omega_\gamma
a_0)^{2\lambda+1}~~,
\end{equation}
 where $S_X$ and $S_\gamma$ are the $S$-factors for
the $X^-$ transfer and normal radiative processes, respectively.
The quantity $a_0$ is the
$X^-$ Bohr radius of $^4$He$_X$ or $^6$Li$_X$, while $p_f$ is the linear momentum
of the outgoing $^7$Li or $^7$Be.  The quantity 
$\omega_\gamma$ is the energy of the emitted photon  of multipole order 
$\lambda$ ($\lambda=1$ for 
electric dipole) in the radiative-capture reactions. 

In the present work, however, we consider more details of the reaction dynamics 
in order  to
better clarify the  differences in these reactions.
First, we note that $^4$He, $^{6,7}$Li, and $^7$Be occupy an s-wave orbit around the $X^-$
particle (assuming the $X^-$ particle to be much heavier than these
nuclei).  At the same time, the $^6$Li nucleus is an $\alpha+d$ cluster system in a
relative s-wave orbit, while the $A=7$ nuclei are $\alpha+t$ and
$\alpha+^3$He cluster systems in relative p-wave orbits. This difference
in the orbital angular momentum will produce a critical difference in
the reaction dynamics between the $^4$He$_X$($d$,$X^-$)$^6$Li reaction and the
$^4$He$_X$($t$,$X^-$)$^7$Li, $^4$He$_X$($^3$He,$X^-$)$^7$Be, and
$^6$Li$_X$($p$,$X^-$)$^7$Be reactions.   In particular,  the latter three
$X^-$ transfer reactions to produce $^7$Li and $^7$Be must involve a  $\Delta
l=1$ angular momentum transfer. This leads to a large hindrance of the
overlap matrix element of the nuclear potential for the $X^-$ transfer
processes.  In the latter three reactions the outgoing $^7$Li and
$^7$Be in the final state must occupy a scattering p-wave orbit from the
$X^-$ particle in order to conserve total angular momentum. Thus, a
realistic quantum mechanical calculation would deduce much smaller
$S_X$-factors than those estimated by~\citet{cyb06}. In this
article, therefore,  we have assumed that the above three reaction processes
are negligible.  

\subsubsection{ $^7$Be$_X$+$p$ Resonant Reaction}

\citet{bir07} have recently suggested that the 
$^7$Be$_X$($p$,$\gamma$)$^8$B$_X$ resonant reaction could occur 
through an atomic excited state of
$^8$B$_X$ with  a  threshold energy of 167 keV.  This channel does  destroy a significant amount of
$^7$Be$_X$ and is included in our present study.

In the previous study~\citep{kus07b}, we have suggested that a
     reaction channel through the $1^+$, $E^*=0.770\pm 0.010$ MeV nuclear excited state
     of $^8$B via $^7$Be$_X$+$p$ $\rightarrow ^8$B$^*$($1^+$, 0.770
     MeV)$_X$ $\rightarrow ^8$B$_X$+$\gamma$ could also destroy
     $^7$Be$_X$.  However, we found that this channel is unimportant
     from the estimate of binding energies in this study.  In~\citet{kus07b}, the binding energies and the eigenstate wave
     functions of the $X$-nuclei were calculated assuming uniform
     finite-size charge distributions of radii $r_0=1.2A^{1/3}$~fm for
     nuclear mass number $A$~\citep{cah81}.  Since the assumed
     charge radii were smaller than the measured charge radii, the
     estimated binding energies were higher than those of this study.
     Although the nuclear resonance channel is not important, we show
     the effect of this channel on $^7$Be$_X$ destruction.

The TRR for resonant radiative capture reactions is given approximately
by 
\begin{equation}
\langle \sigma v \rangle_{\rm R} = \hbar^2\left(\frac{2\pi}{\mu k_BT}\right)^{3/2} \omega_\gamma \exp(-E/k_BT),
\label{eq3}
\end{equation}
\begin{equation}
\omega_\gamma=\frac{2I+1}{\left(2I_1+1\right)\left(2I_2+1\right)} \frac{\Gamma_l \Gamma_\gamma}{\Gamma_{\rm tot}},
\label{eq4}
\end{equation}
where $\hbar$ is Planck's constant, $E$ is the resonance
energy, and $I_1$, $I_2$, and $I$ are the spins of the projectile, the
target, and the resonance.  The quantity $\Gamma_l$ is the
particle width for decay into two charged particles of relative
angular momentum $l$, while  $\Gamma_\gamma$ is the gamma decay width, 
and $\Gamma_{\rm tot}$ is the total width of the
resonant state. 

$\Gamma_l$ is approximately written~\citep{boy07} as
\begin{eqnarray}
\Gamma_l &\approx& \frac{3\hbar}{R} \left(\frac{2}{AM_{\rm u}} \right)^{1/2} \theta_l^2 E_c^{1/2} \nonumber \\
&&\times \exp [bE^{-1/2}+1.05\left(ARz_1Z_2\right)^{1/2} \nonumber \\
&&-7.62\left(l+\frac{1}{2}\right)^2 (ARz_1Z_2)^{-1/2}],
\label{eq5}
\end{eqnarray}
where $R=1.4(A_1^{1/3}+A_2^{1/3}) \times 10^{-13}$~cm is the interaction
radius with $A_1$ and $A_2$ the atomic weights of the interacting
particles.  The symbol $A$ denotes  the reduced atomic weight $A=A_1A_2/(A_1+A_2)$,
and $M_{\rm u}$
is the atomic mass unit,  while $\theta_l^2$ is the dimensionless reduced width. The height of the
Coulomb barrier is 
\begin{equation}
E_c=1.44({\rm MeV~fm}) \frac{z_1 Z_2}{R},
\label{eq6}
\end{equation}
while the usual Sommerfeld parameter $b$ is given by  $b=31.28 z_1 Z_2
A^{1/2}($keV$^{1/2}$).  Correcting the charge, reduced mass, and energy
for the nuclides with an embedded $X^-$ particle, we obtain the partial
width of the $^8$B$^\ast$(0.77 MeV)$_X \rightarrow ^7$Be$_X$+$p$ reaction,
\begin{equation}
\Gamma_{1,X} \approx 1.7 \times 10^6~{\rm eV} \exp\left(-\frac{93.9}{(E_{\rm th}/{\rm keV})^{1/2}}\right),
\label{eq7}
\end{equation}
where $E_{\rm th}$ is the energy level of the nuclear resonant state with
respect to the $^7$Be$_X$ plus $p$ exit channel.  

From the conjugate analog
state of $^8$Li ($1^+$, 0.9809 MeV) we deduce that the $^8$B (0.770 MeV) first
excited state has spin and parity $I^\pi=1^+$ and that the resonant
reaction proceeds through a p-wave ($l=1$).  In Eq.~(\ref{eq7}) we also
adopted $\theta_1^2=0.82$ from the TRR of this resonance used
in standard BBN.  Applying this proton width [Eq.~(\ref{eq7})] and
$\Gamma_\gamma=25 \pm 4$~meV~\citep{ajz88} to Eqs.~(\ref{eq3})
and~(\ref{eq4}), we can estimate the resonant TRR for $^7$Be$_X$+$p
\rightarrow ^8$B$^\ast$($1^+$, 0.770 MeV)$_X \rightarrow
^8$B$_X$+$\gamma$ as a function of $E$.  When $E \approx 0$,
$\Gamma_{1,X}$ vanishes, thus $\omega_\gamma \rightarrow 0$ and the TRR
also vanishes.  As the resonance energy $E$ increases, $\Gamma_{1,X}$
also gradually increases and eventually becomes larger than
$\Gamma_\gamma$.  Thus, $\omega_\gamma$ converges to a constant value
$\omega_\gamma=(2I+1)/((2I_1+1)(2I_2+1))\Gamma_\gamma$ in
Eq.~(\ref{eq4}).  However, the exponential factor exp($-E/k_BT$) in
Eq.~(\ref{eq3}) strongly regulates the TRR for larger $E$.  In this
manner there is a most effective resonance energy $E_{\rm eff}$ which
maximizes the TRR.  We found $E_{\rm eff} \approx 30$~keV
numerically. Hence, this resonant reaction could be an important, possibly dominant, new means to destroy
$^7$Be$_X$.  The present study, however, indicates that is not the case.

In the previous study~\citep{kus07b} the calculated binding energies of the $X^-$ particle in
$^7$Be$_X$ and $^8$B$_X$ were respectively 1.488 MeV and 2.121 MeV.
If we adopt these values for the energy levels of the
nuclear excited states of $^8$B$_X$, this $1^+$ state of $^8$B$_X$
becomes located  near the particle threshold for the $^7$Be$_X$+$p$ 
channel.  In this case the $^7$Be$_X$($p$,$\gamma$)$^8$B$_X$ reaction can
proceed through a zero-energy resonance of $^8$B$^\ast_X$ at $E\approx 0$~MeV, where $E$ is the center-of-mass energy between the  $^7$Be$_X$ and the
proton  in
the entrance channel.  We note, however, that the estimated binding energies of $X$-nuclei,
depend upon several assumptions such as the
adopted charge distribution of the $X^-$ embedded nucleus~\citep{pos07,bir07}.  The $A$=6, 7, and 8 nuclear systems are
typical clustering nuclei for which even a small change of the relative wave
function between the composite nuclei can affect significantly the
radiative capture cross sections at astrophysical energies as well as  their static
electromagnetic properties~\citep{kaj86,kaj88}.  We therefore adopt a large uncertainty in
 the $1^+$ resonance energy, $E$, from the $^7$Be$_X$+$p$ separation
threshold in~\citet{kus07b}.

To check the nuclear flow to the higher mass region we added the $^8$Be$_X$+$p$ $\rightarrow ^9$B$_X^{\ast{\rm a}}
\rightarrow ^9$B$_X$+$\gamma$ reaction, where $^9$B$_X^{\ast{\rm a}}$
indicates that the reaction occurs through an atomic excited state
of $^9$B.  However this reaction was now found to be unimportant because its threshold energy is
relatively large (see Table \ref{tab2}).

\subsection{Reaction Network}

\begin{figure}
\plotone{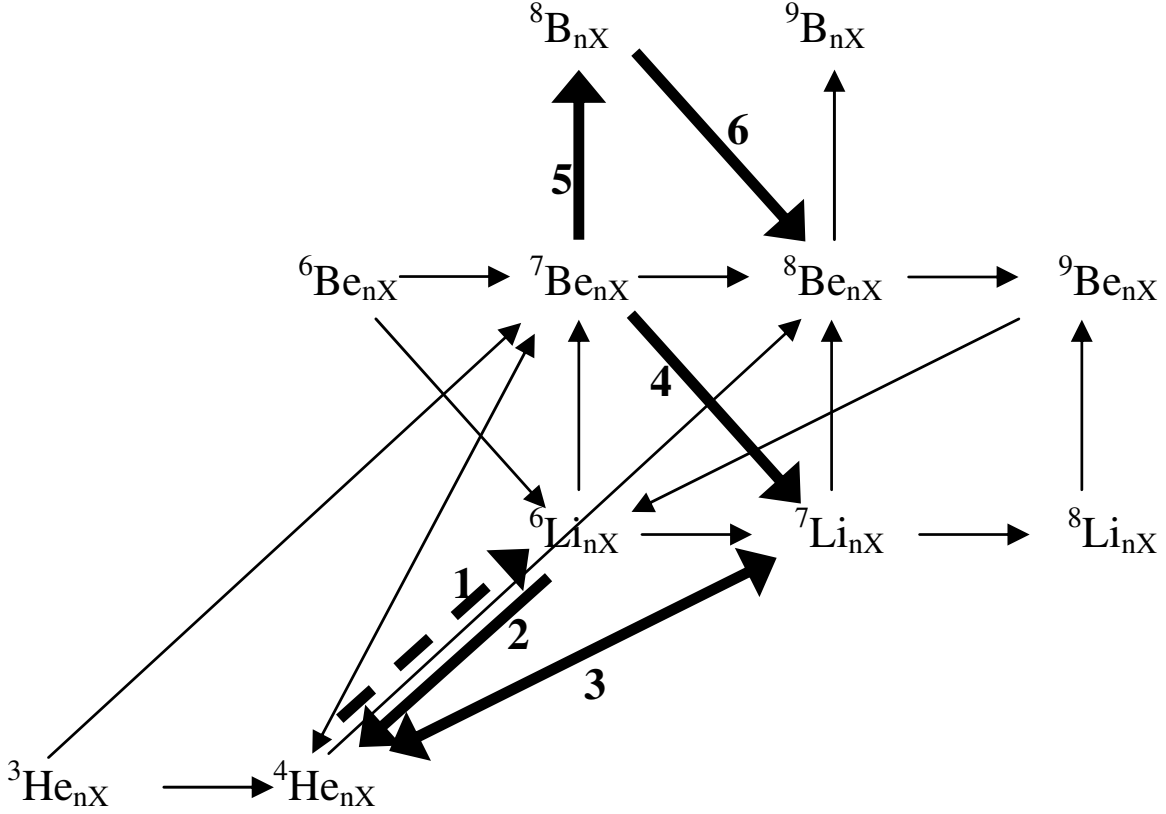}
\caption{Reaction pathways (solid arrows along the directions of positive
 $Q$-values) that ultimately occur for the nuclides with an embedded
 $X^-$ particle.  The integer ^^ ^^ $n$'' can be 0 or 1.  Thick solid
 arrows indicate the most important reactions which contribute in
 synthesizing or destroying $^6$Li$_X$, $^7$Li$_X$, and
 $^7$Be$_X$.  A thick dashed arrow indicates the $X^-$ transfer reaction 
 $^4$He$_X$($d$,$X^-$)$^6$Li.  Numbers attached indicate: 1.
 $^4$He$_X$($d$,$X^-$)$^6$Li; 2. $^6$Li$_X$($p$,$^3$He)$^4$He$_X$; 3. $^4$He$_X$($t$,$\gamma$)$^7$Li$_X$ \& $^7$Li$_X$($p$,$\alpha$)$^4$He$_X$;
 4. $^7$Be$_X$(,$X^0$)$^7$Li; 5. $^7$Be$_X$($p$,$\gamma$)$^8$B$_X$;
 6. $^8$B$_X$(,$e^+ \nu_e$)$^8$Be$_X$\label{fig1}}
\end{figure}

We show in Fig.~\ref{fig1} the reaction network involving the
nuclides with an embedded $X^-$ particle relevant to the present paper.
Nuclear reactions denoted by the thick solid arrows were found to be
especially important for the production or destruction of the $A$=6 and
7 nuclides both in the literature and in our present study.  \citet{pos07}
originally proposed the main production process for $^6$Li to be the  
$^4$He$_X$($d$,$X^-$)$^6$Li reaction.   Similar $X^-$ transfer reactions
$^4$He$_X$($t$,$X^-$)$^7$Li, $^4$He$_X$($^3$He,$X^-$)$^7$Be,
$^6$Li$_X$($p$,$X^-$)$^7$Be were discussed by~\citet{cyb06} as
possible additional production processes of $A$=7 nuclides.  Here we additionally considered the
destruction processes $^6$Li$_X$($p$,$^3$He)$^4$He$_X$,
$^7$Li$_X$($p$,$\alpha$)$^4$He$_X$, and $^7$Be$_X$($d$,$p
\alpha$)$^4$He$_X$.  We find, however, that these latter three reactions
produce small effect on the BBN results and, as noted above, the latter
three $X^-$ transfer reactions would have little effect on BBN results when
$\Delta l=1$ hindrance is taken into account.  

\citet{bir07}
have proposed the likely important destruction process $^7$Be$_X$+$p$ $
\rightarrow ^8$B$_X^{\ast{\rm a}} \rightarrow ^8$B$_X$+$\gamma$, where
$^8$B$_X^{\ast{\rm a}}$ is an atomic excited state through which
the radiative capture occurs.  They also proposed a charged weak-boson
exchange reaction $^7$Be$_X \rightarrow ^7$Li+$X^0$ followed by
$^7$Li($p$,$\alpha$)$^4$He and $^7$Li($X^-$,$\gamma$)$^7$Li$_X$($p$,$\alpha$)$^4$He$_X$ to eventually destroy the $A$=7 nuclides.  Other reactions can also contribute
to synthesize or destroy $^6$Li$_{nX}$ and $^7$Li$_{nX}$, $^7$Be$_{nX}$
($n$=0 or 1) and the heavier nuclei.  Our network code includes
many reactions with nuclei up to Carbon isotopes.  Table \ref{tab2}
summarizes our adopted nuclear reaction rates.

\subsection{Kinetic Rate Equations}\label{sec25}

When dealing with the kinetic and chemical equilibrium associated with
$X^-$ particles, it is necessary to consider the thermodynamics
associated with the binding of $X^-$ particles. This is because it will be
important to know precisely when during BBN they become bound to nuclei,
and what their distribution over the BBN nuclei is expected to
be~\citep{koh07}.  We have put both recombination and ionization
processes of $X^-$ particles into our BBN network code and have dynamically
solved the associated  set of rate equations [as in~\citet{koh07}] to find when the $X$-nuclei decoupled from the cosmic
expansion.  Denoting a specific isotope of an element by $(N,Z)$, its
abundance as $n(N,Z)$, and the corresponding quantities for the isotopes
with an embedded $X^-$ as $(N,Z)_X$ and $n(N,Z)_X$, the
capture-reionization expressions for $(N,Z)$ and $(N,Z)_X$ in an expanding 
universe are
\begin{eqnarray}
&&\frac{\partial n(N,Z)}{\partial t} + 3Hn(N,Z) \nonumber \\
&&~~~ =\left[\frac{\partial n(N,Z)}{\partial t}\right]_{\rm creation} - \left[\frac{\partial n(N,Z)}{\partial t}\right]_{\rm destruction} \nonumber \\
&&~~~~~~~- \left[\frac{\partial n(N,Z)_X}{\partial t}\right]_{\rm capture},
\label{eq8}
\end{eqnarray}
and
\begin{eqnarray}
&&\frac{\partial n(N,Z)_X}{\partial t} + 3Hn(N,Z)_X \nonumber \\
&&~~~ =\left[\frac{\partial n(N,Z)_X}{\partial t}\right]_{\rm creation} - \left[\frac{\partial n(N,Z)_X}{\partial t}\right]_{\rm destruction} \nonumber \\
&&~~~~~~~+ \left[\frac{\partial n(N,Z)_X}{\partial t}\right]_{\rm capture}.
\label{eq9}
\end{eqnarray}
Here,  the subscript ^^ ^^ creation'' refers to nuclear reactions that
make $(N,Z)$ and ^^ ^^ destruction'' to nuclear reactions that destroy
it, including $\beta$-decay for unstable nuclides.  The quantity  $H$ is the Hubble
expansion rate.  Detailed balance of the $X$-capture reaction $(N,Z) +
X^- \rightarrow (N,Z)_X + \gamma$ and its inverse permits~\citep{koh07} writing the capture process as
\begin{equation}
\left[\frac{\partial n(N,Z)_X}{\partial t}\right]_{\rm capture} \approx \langle \sigma_{\rm r} v \rangle \left(n_X n(N,Z) - n(N,Z)_X \tilde{n}_\gamma\right),
\label{eq10}
\end{equation}
where $\sigma_{\rm r}$ is the recombination cross section, $n_X$ is the
abundance of $X^-$ particles, and $\tilde{n}_\gamma$ is the number of
photons in excess of $E_{\rm
bind}$, the $X^-$ binding energy to $(N,Z)$, i.e.,
\begin{equation}
\tilde{n}_\gamma = n_\gamma \left(\frac{\pi^2}{2 \zeta(3)}\right)\left(\frac{m_{(N,Z)}}{2 \pi T}\right)^{3/2} \exp\left(-\frac{E_{\rm bind}}{k_B T}\right),
\label{eq11}
\end{equation}
and
\begin{equation}
n_\gamma = \left(\frac{2\zeta(3)}{\pi^2}\right)T^3.
\label{eq12}
\end{equation}
Note that since $E_{\rm bind}$ is small, the equilibrium will favor
unbound $X^-$ particles until low $T$.  Since the mass of the $X^-$ particle is assumed to be $\gtrsim 50$~GeV, the
reduced mass for the $X^-+A(N,Z)$ system can be approximated as $\mu_X
\equiv m_A m_X/(m_A+m_X) \approx m_A$,.  This leads to the following TRR for the first
recombination process $A$($X^-$,$\gamma$)$A_X$~\citep{koh07}
\begin{eqnarray}
\langle \sigma_{\rm r} v \rangle_X & \approx & \frac{2^9 \pi \alpha Z^2 \left(2\pi \right)^{1/2}}{3\exp(4.0)} \frac{E_{\rm bind}}{\mu_X^2 \left(\mu_X T\right)^{1/2}}\nonumber \\
& \propto & Z^2 E_{\rm bind} m_A^{-2.5} \sim Z^3 m_A^{-2.5}.
\label{eq13}
\end{eqnarray}
where $\alpha$ is the fine
structure constant.  This expression is almost independent of $m_X$ but
increases with $Z$.  We do, however, obtain 
a different mass-dependence in the
TRR for the second recombination process $A_X$($X^-$,$\gamma$)$A_{XX}$:
\begin{eqnarray}
\langle \sigma_{\rm r} v \rangle_{XX} & \approx & \frac{2^9 \pi \alpha \left(Z-1\right)^2 \left(2\pi \right)^{1/2}}{3\exp(4.0)} \frac{E_{\rm bind}}{\mu_{XX}^2 \left(\mu_{XX} T\right)^{1/2}}\nonumber \\
& \propto & (Z-1)^2E_{\rm bind}m_X^{-2.5} \sim (Z-1)^3 m_X^{-2.5}. \nonumber \\
\label{eq14}
\end{eqnarray}
Here we obtain a mass dependence because $\mu_{XX} \equiv m_{AX} m_X/(m_{AX}+m_X) \approx m_X/2$.  Since $m_X$
is typically much larger than the mass of the light nuclei $m_X \gg
m_A$, the rate for the second or higher-order recombination process is
hindered.

\section{RESULTS}\label{sec3}

\subsection{BBN Calculation Result}

\begin{figure}
\epsscale{0.80}
\plotone{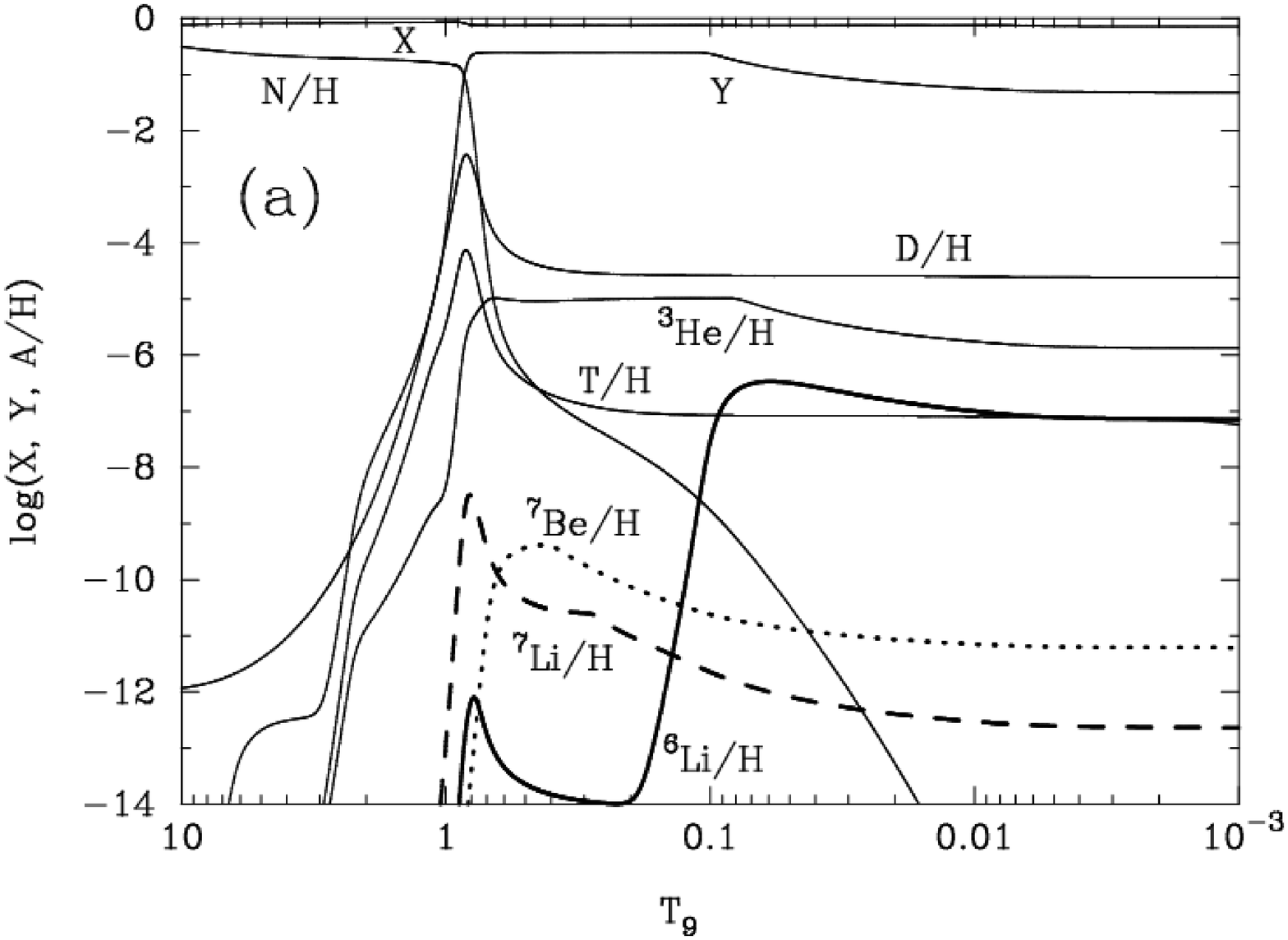}
\plotone{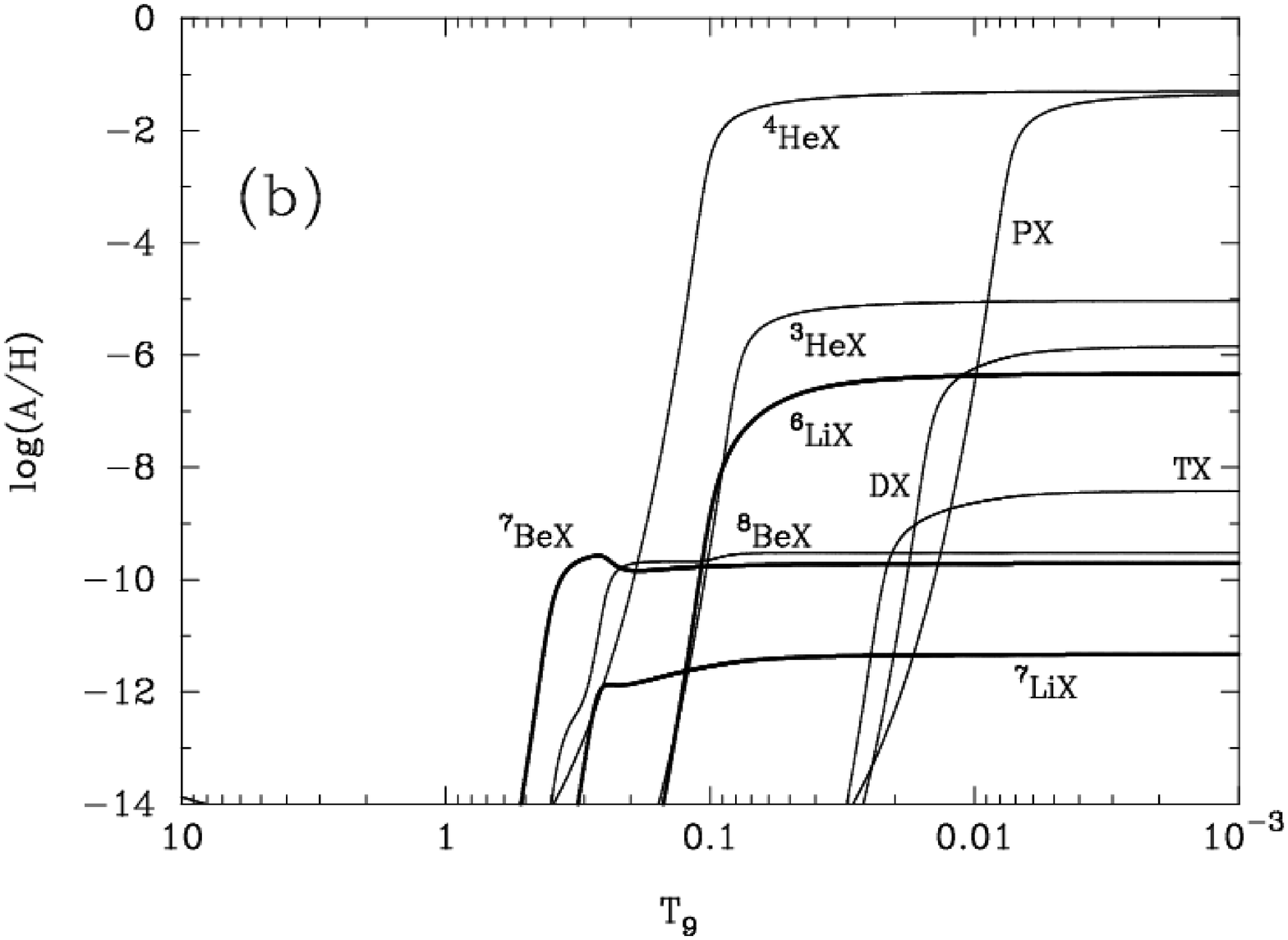}
\caption{Calculated abundances of normal nuclei (a) and $X$-nuclei (b)
 as a function of $T_9$.  For this figure we have taken the abundance of negatively charged
 particles $X^-$ to be $Y_X=n_X/n_b=0.1$, and its lifetime is taken to
 be long $\tau_X=\infty$.  We
 utilize the $X^-$ reaction rates as described in the text.\label{fig2}}
\end{figure}

Figures \ref{fig2}a and \ref{fig2}b illustrate the results of a BBN calculation 
in which the $X^-$ abundance is
taken to be 10\% of the total baryon number,  i.e. $Y_X=n_X/n_b=0.1$, where $n_X$ is the number density of the $X^-$ particles and
$n_b$ is the averaged universal baryon-number density.
Results shown in Fig.~\ref{fig2}a  are for normal nuclei while Fig.~\ref{fig2}b
shows those for the
$X$-nuclei,   Note that we
adopt the rate of~\citet{ham07} for $^4$He$_X$($d$,$X^-$), and the reactions $^4$He$_X$($t$,$X^-$),
$^4$He$_X$($^3$He,$X^-$), and $^6$Li$_X$($p$,$X^-$) are taken to be negligible.  

The
abundances for the normal nuclei $^6$Li, $^7$Li and $^7$Be in the interval $T_9 \sim
0.5-0.2$ are seen to be close to their usual  BBN values.  This is
because at higher temperatures the nuclear statistical equilibrium does
not particularly favor the production of the weakly bound $X$-nuclei.
Thus, their effect on the abundance is minimal.  There are some changes, however, 
once the $X$-nuclei appear.  Because of the larger binding energies, the $X^-$
particles bind first to the heaviest nuclides, such as $^7$Li and $^7$Be
produced in normal BBN (first recombination of the $X^-$ particles).
These recombinations occur at around $T_9=0.3$ (for $^7$Li) and $T_9=0.5$
(for $^7$Be), respectively.   An increase in the $^7$Be$_X$ abundance by the
recombination can clearly be seen in Fig.\ \ref{fig2}b.  Somewhat later, at
around $T_9=0.1$, the $X^-$ particles are captured onto $^4$He, as can
be seen in Fig.\ \ref{fig2}b.  Then a new round of nucleosynthesis of
the $X$-nuclei, involving the reaction $^4$He$_X$($d$,$X^-$), produces
normal nuclei $^6$Li (in Fig.\ \ref{fig2}a) as well as $^6$Li$_X$ (in
Fig.\ \ref{fig2}b).  However, the most notable feature of these results is that the
$^6$Li is not easily destroyed by the $^6$Li($p$,$\alpha$)$^3$He
reaction, which destroys nearly all of the $^6$Li produced in standard BBN.  This is  because the $X^-$ transfer reaction restores the charge of the
normal nucleus $^6$Li, which has a Coulomb barrier which is too
high at these temperatures for
its destruction.  Thus, the large abundance ratio of mass 6 to mass 7 is
preserved~\citep{pos07}.  In addition, however, 
$^7$Be is destroyed by the nuclear reactions that occur after the recombination
($T_9 \sim 0.3$), mainly
$^7$Be($X^-$,$\gamma$)$^7$Be$_X$($p$,$\gamma$)$^8$B$_X$.  This is
explained in detail below.

The calculated BBN abundances of the mass 6 and 7
nuclides and other light nuclei depend strongly on the $X^-$ abundance.  The
above discussion applies only to the case of relatively abundant $X^-$
particles.  In order to study the sensitivity of the $^6$Li and $^7$Li
($\equiv ^7$Li+$^7$Be) abundances to the $X^-$ abundance, $n_X$, we carried out
a series of  BBN calculations in which $n_X$  was varied over a wide range.  

\begin{figure}
\epsscale{1.}
\plotone{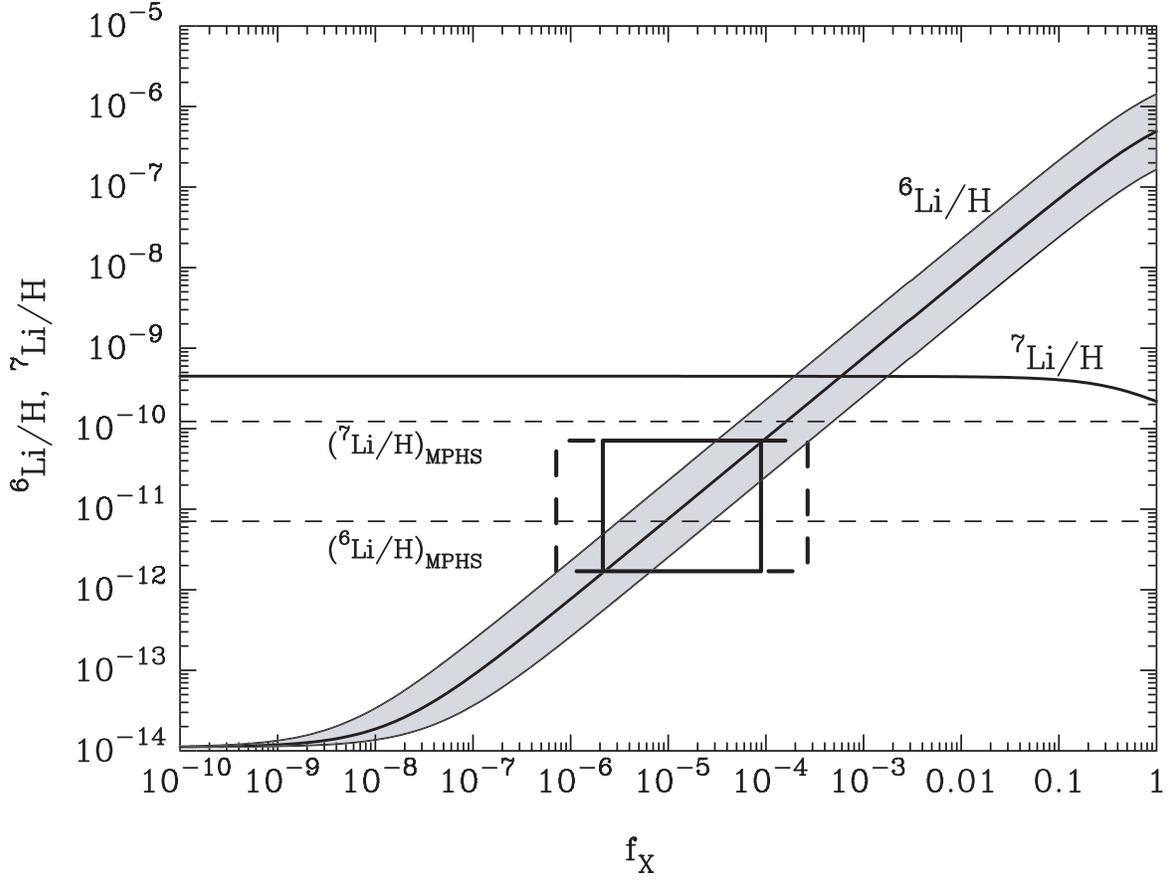}
\caption{Calculated abundances of $^6$Li/H and $^7$Li/H as a function of the initial $X^-$
 abundance parameter $f_X$ defined in the text.  The gray band bounded
 by curves of $^6$Li show the uncertainty from the rate for
 $X^-$ transfer reaction.  The dashed lines indicate the mean values
 observed in MPHSs.  The dashed and solid boxes indicate the range of
 $f_X$ consistent with our adopted limits on the abundance of $^6$Li/H
 observed in MPHSs.\label{fig3}}
\end{figure}

Figure~\ref{fig3} shows the calculated abundances of the mass 6 and 7 nuclides.
These are plotted as $^6$Li/H (solid curves of positive slope) and $^7$Li/H
(horizontal solid
curve), as a function of the initial $X^-$ number fraction $f_X$ relative to the cold dark matter (CDM) abundance, which we define as 
\begin{equation}
 f_X \equiv \frac{Y_X}{0.09}  =  \frac{n_X}{n_{\rm CDM}}  \biggl(\frac{50~{\rm GeV}}{m_{\rm CDM}}\biggr)~~.
\label{eq15}
\end{equation}
This definition follows if we take  number density  of CDM particles as that inferred
 from the CDM closure content ($\Omega_{\rm CDM} = 0.2$) deduced from
 the WMAP data.  For this closure parameter, the  CDM number density, for any value of
 mass $m_{\rm CDM}$  of the CDM particle,  can be written $n_{\rm
CDM} = 0.09~n_b \times (50~{\rm GeV}/m_{\rm CDM})$. So, our definition of $f_X$ is for  a fiducial  CDM  mass of  50 GeV and is easily scalable to other CDM masses.

 In the calculations shown in Fig.\ \ref{fig3} we have assumed that
the $X^-$ particle has a mean lifetime much longer than  the
typical time scale for BBN in the presence of $X^-$ particles, 
i.e $\tau_X \gg 5$~hours.  Below we also consider the case of a mean lifetime which is shorter than 5 hours.
In this figure we have taken into  account the theoretical uncertainty in the $X^-$
transfer reaction cross sections.  For $^6$Li the upper curves
correspond to the yields with our assumed rate for the
$^4$He$_X(d,X^-)^6$Li reaction \citep{ham07}
multiplied by a factor of 3, and the lower curves correspond to that
same rate divided by a factor of 3. The middle curve for $^6$Li
corresponds to our best guess for that crucial rate, i.e., the rate of \citet{ham07}.  As noted
above,~\citet{bir07} suggested that the recombination of $^7$Be and
$X^-$ together with the $^7$Be$_X(p,\gamma)^8$B$_X$ reaction taking
account of the resonant contribution from both reactions would destroy a considerable amount of $^7$Be; this process was included in all of our calculations.

It is clear from  Fig.~\ref{fig3} that the $^6$Li abundance increases
monotonically with increasing $f_X$.  This is a consequence of the
fact that $^6$Li is mainly produced by the $^4$He$_X$($d$,$X^-$)$^6$Li reaction as 
proposed in~\citet{pos07} for almost entire range of $f_X$-values.  For $f_X
\lesssim 10^{-9}$, however,  the standard BBN processes
$^4$He($d$,$\gamma$)$^6$Li  and $^3$He($t$,$\gamma$)$^6$Li are the main
reactions to make $^6$Li~\citep{fuk90,smi93}.  In the  region of $f_X
\gtrsim 0.1$, however, a departure from
the linear increase due to
$^6$Li($X^-$,$\gamma$)$^6$Li$_X$($p$,$^3$He)$^4$He$_X$ is observed.

 During normal BBN, at the 
baryon to photon ratio deduced from CMB measurement by WMAP,
 $^7$Li is produced mainly as $^7$Be.  We find that the
 $^7$Be (and $^7$Li) produced during  the standard BBN epoch 
captures $X^-$ particles to form $^7$Be$_X$ (and $^7$Li$_X$) during the
first recombination.   The  $^7$Be is then destroyed by the 
$^7$Be$_X$($p$,$\gamma$)$^8$B$_X$ reaction.  In the parameter region of lower $X^-$
abundance $f_X \lesssim 0.1$, $^7$Be and $^7$Li are produced in
the standard BBN by the  $^4$He($t$,$\gamma$)$^7$Li and
$^4$He($^3$He,$\gamma$)$^7$Be reactions.  They are not, however, 
 destroyed and remain almost unchanged.

\subsection{Constraints on the Primordial $^6$Li Abundance}\label{sec32}
Constraints on the primordial $^6$Li abundance must be  inferred
from the observed plateau as a function of metallicity on MPHSs.  
The actual primordial abundance of  $^6$Li  could be higher than the recently detected high
plateau-abundance~\citep{asp06} of $^6$Li (lower horizontal dashed curve
in Fig.\ \ref{fig3}).  This is because stellar processing could have depleted an initial surface 
abundance.  
It is expected in models~\citep{pin02} of
stellar structure and evolution that both $^6$Li and $^7$Li abundances
decrease because materials on the stellar surface might be
convected to regions of sufficiently high temperature that 
 the fragile $^6$Li and $^7$Li are partially
 destroyed~\citep[e.g.][]{lam04,ric05}.  Therefore, the plateau level
for the observed abundances of $^6$Li/H and $^7$Li/H in MPHSs should be
considered a lower limit to the primordial abundance.  For this lower limit we take the 3~$\sigma$ lower limit to the mean
plateau value times a factor of $1/3$ for systematic uncertainties  giving 
$^6$Li/H $\ge 1.7 \times 10^{-12}$.  We include this additional factor
because there may be additional systematic
uncertainties due to the sensitivity of the inferred $^6$Li abundance to
the model atmosphere employed~\citep{cay07}.

The upper limit is more difficult to estimate.  Because  $^6$Li could be more easily destroyed in stars than $^7$Li, its upper limit should be higher than the upper limit to the $^7$Li abundance.
Even so, there are constraints on the degree of stellar processing for both $^7$Li and $^6$Li
from the limits on the dispersion of the plateau.  A large degree of
stellar destruction would be sensitive to the varying degrees of meridional circulation in the stars~\citep{pin02}
and hence should produce a large dispersion in observed abundances.
However, the fact that the observed dispersion in $^7$Li  is greater than that observed for $^6$Li 
(albeit on a limited data set) suggests that the destruction may not be
very significant~\citep{pin02}.  
For this reason we adopt a conservative upper limit of a factor of 10
above the mean plateau value giving
$^6$Li/H $\le 7.1\times 10^{-11}$.

\subsection{Observational Constraints on the $X^-$ Abundance}

\subsubsection{Case of Longer Mean Life of $X^-$ Decay}

\begin{figure}
\plotone{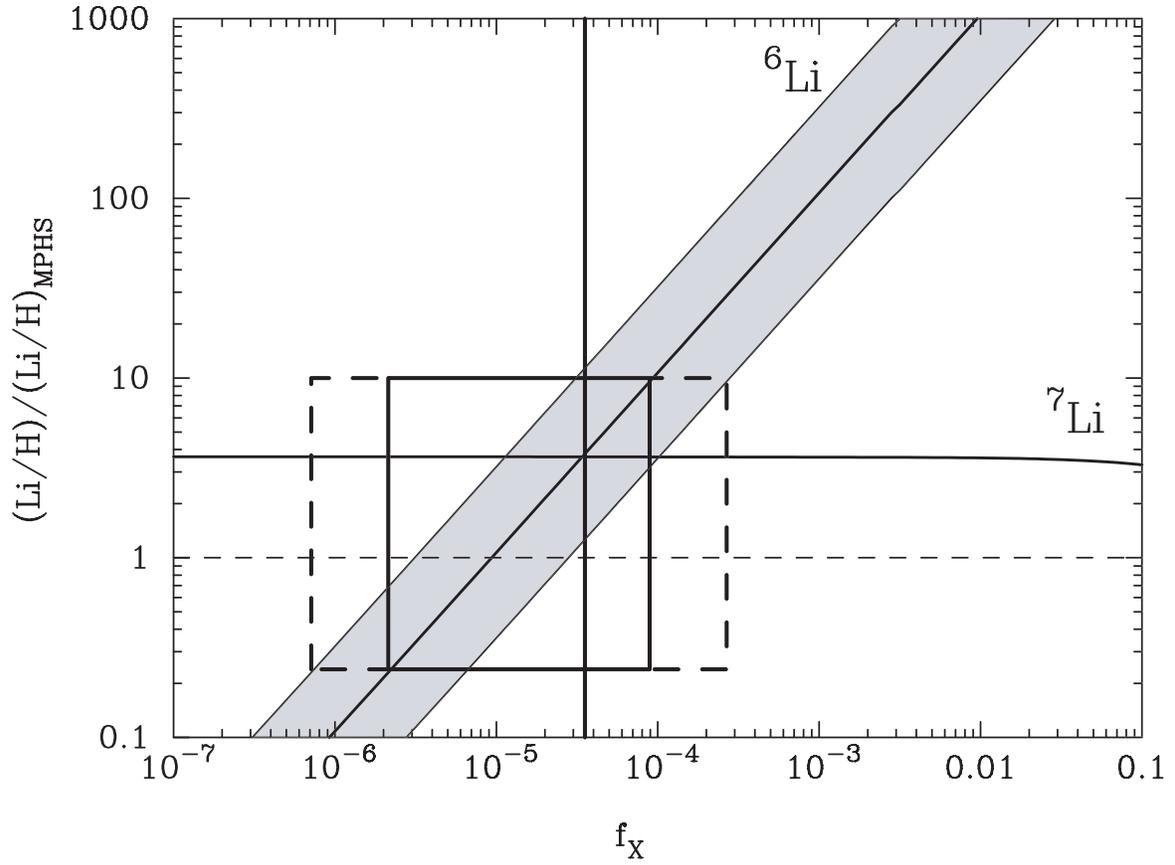}
\caption{Calculated abundances of $^6$Li/H and $^7$Li/H normalized
 to the mean value observed in MPHSs as a
 function of $f_X$.  The solid (dashed) boxes are similar to
 those of Fig.\ \ref{fig3}.  The vertical solid line
 corresponds to the lowest abundance of $X^-$ ($f_X$) which leads to a
 larger enhancement of $^6$Li than $^7$Li.\label{fig4}}
\end{figure}

We consider several constraints on the $X^-$ abundance from the
observed Li isotopic abundances.  We first discuss the case of
a longer mean life for the $X^-$ particle ($\tau_X \gtrsim~5$ hours).
In this case, the yields of BBN are not strongly affected by the value
of $\tau_X$.  Our adopted  lower limit to the $^6$Li/H ratio is then satisfied by the
calculated abundance of $^6$Li/H (thick solid curve in Figs.~\ref{fig3}
and \ref{fig4}) for $ f_X \gtrsim 2 \times
10^{-6}$. 

In order to include the uncertain depletion effect of fragile lithium in stellar
atmospheres, we display in Fig.\ \ref{fig4} the calculated primordial
$^6$Li and $^7$Li abundances relative to the mean plateau levels in MPHSs, i.e. ($^A$Li/H)/ ($^A$Li/H)$_{\rm MPHS}$.  The
gray regions enclose our adopted uncertainty due to the  $X^-$ transfer reaction
cross sections as in Fig.\ \ref{fig3}.  Here and below, we take, as our
recommended network, the reaction rate of $^4$He$_X$($d$,$X^-$)$^6$Li
from~\citet{ham07}.

The parameter
regions of initial $X^-$ abundance, $2 \times 10^{-6} \lesssim f_X
\lesssim 1 \times 10^{-4}$ for $^6$Li and $f_X \lesssim 1$ for $^7$Li,
respectively, are plausible regions in this context.  In addition, since
$^6$Li is more easily destroyed in proton burning than $^7$Li, the
primordial abundances should satisfy the inequality
($^7$Li/H)/($^7$Li/H)$_{\rm MPHS} \lesssim$ ($^6$Li/H)/($^6$Li/H)$_{\rm
MPHS}$.  From Fig.\ \ref{fig4} we thus find the concordant parameter
region of
\begin{eqnarray}
4 \times 10^{-5} \lesssim f_X \lesssim 9 \times 10^{-5}, \nonumber \\
({\rm i.e.}~4 \times 10^{-6} \lesssim Y_X \lesssim 8 \times 10^{-6}),
\label{eq17}
\end{eqnarray}
which is bounded by the vertical solid line on Fig.~\ref{fig4}.  

Here, we used Eq.~(\ref{eq15}) to convert the inferred limits on $f_X$
into the limits for $Y_X$.  We then deduce from Eq.~(\ref{eq17}) and Fig.\
\ref{fig4} a possible depletion factor, $d$($^A$Li), of the primordial
abundances of $^6$Li and $^7$Li in Population II metal-poor halo stars of,
\begin{equation}
d(^6{\rm Li}) \lesssim 10,
\label{eq18}
\end{equation}
\begin{equation}
d(^7{\rm Li}) \lesssim 4,
\label{eq19}
\end{equation}
where the upper limit to $^6$Li depletion comes from our adopted upper limit to the
primordial $^6$Li abundance.  We note that these depletion factors are
consistent with the known thermonuclear reaction rates for $p+^6$Li and $p+^7$Li, assuming partial depletion of lithium in the solar atmosphere; the former is about a factor of 80 larger than the latter.  
We also point out that in our present scenario for BBN
including the $X^-$ particles, their abundance parameter region,
Eq.~(\ref{eq17}) can be a solution to the $^6$Li abundance discrepancy between the standard BBN prediction and the
observations of MPHSs, while still satisfying the independent abundance
constraint on the primordial $^7$Li abundance.  It is to be noted, however, that
there still remains a possible controversy that the depletion factor
of $d$($^7$Li)$=3-4$ may be too large to accommodate  the observed small dispersion in
the plateau abundance level detected in MPHSs~\citep{pin02}.

\subsubsection{Case of Shorter Mean Life of $X^-$ Decay}

To study the effects of $X^-$ decay  we introduce the 
lifetime $\tau_X$ of the $X^-$ particle.  Then there are  two parameters
$Y_X$ and $\tau_X$.  Even in the case when the mean life of the $X^-$ particle is nearly
equal to or slightly shorter than five hours ($\tau_X \lesssim 2
\times 10^4$~s) the recombination of $^7$Be and $X^-$ particles 
still enriches the $^7$Be$_X$ abundance (Fig.\ \ref{fig2}b).  
This is because the recombination of $^7$Be and $X^-$ particles
occurs at earlier times when the cosmic temperature is $T_9 \sim 0.5$.
However, in this case the recombination of $^4$He  cannot
produce abundant $^4$He$_X$.  This is because the recombination of $^4$He occurs at 
at a lower temperature $T_9 \sim 0.1$ when the
cosmic time is of the same order of $\tau_X \sim 10^4$~s and is sensitive to the decay.
Thus, the $^6$Li production is reduced because the 
$^4$He$_X$($d$,$X^-$)$^6$Li reaction is strongly hindered.  

   In this case, however, the resonant reaction
process $^7$Be$_X$+$p \rightarrow ^8$B$_X^{\ast {\rm a}} \rightarrow
^8$B$_X$+$\gamma$~\citep{bir07} can more effectively destroy
$^7$Be$_X$.  We assumed that $^8$B$_X$ inter-converts to $^8$Be$_X$ by $\beta$-decay with a rate of $^8$B $\beta$-decay
multiplied by the correction term $(Q_X/Q)^5$, where $Q$ and $Q_X$ are
the $Q$-values of standard $\beta$-decay and that of $\beta$-decay for
$X$-nuclei.  We adopt here the~\citet{ham07} rate for
$^4$He$_X$($d$,$X^-$)$^6$Li.  

\begin{figure}
\plotone{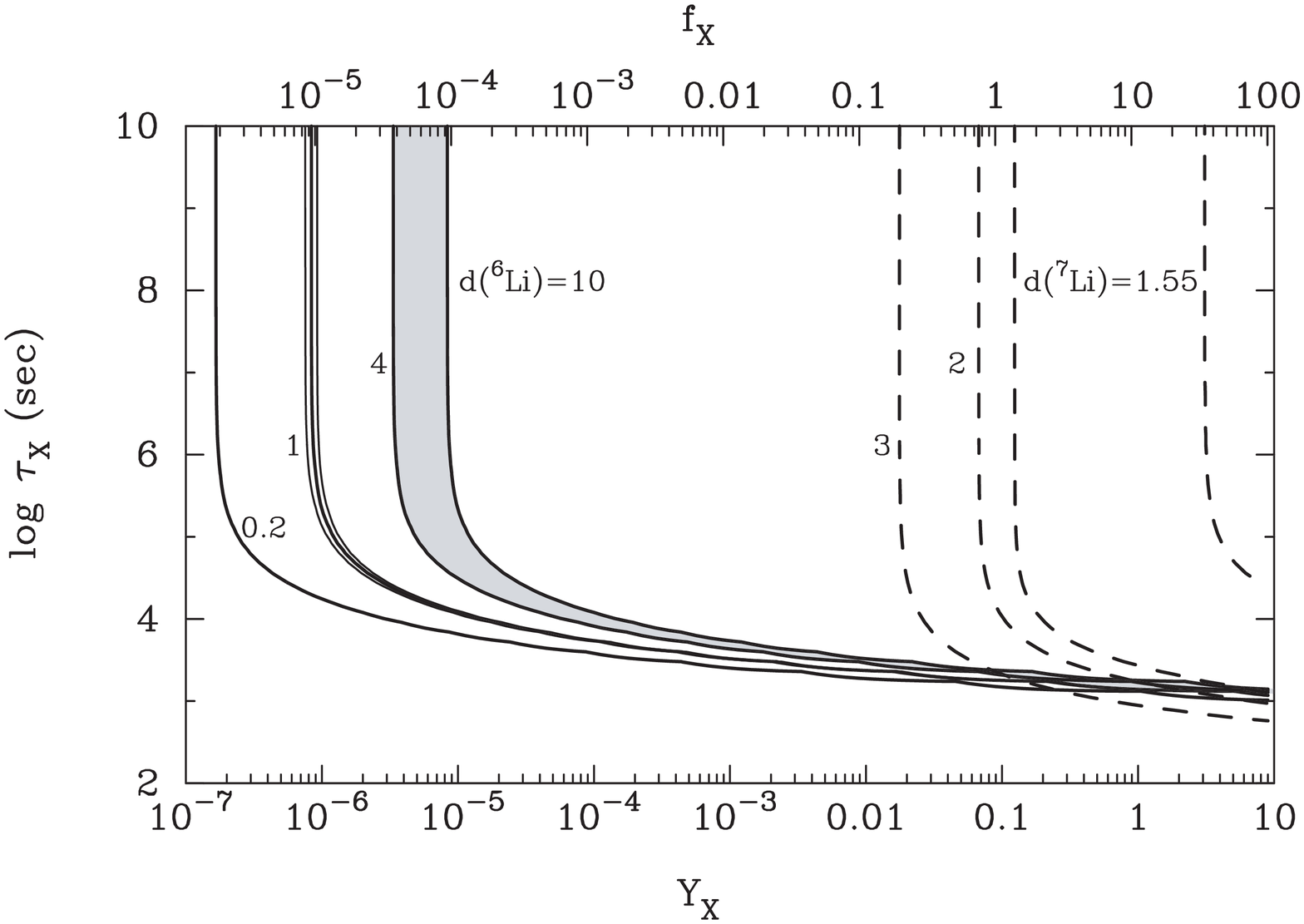}
\caption{Contours of constant lithium abundance relative to the value
 observed in MPHSs, i.e., $d$($^6$Li) = $^6$Li$^{\rm Calc}$/$^6$Li$^{\rm Obs}$ (solid curves) and $d$($^7$Li) = $^7$Li$^{\rm Calc}$/$^7$Li$^{\rm Obs}$ (dashed curves).  The adopted abundances are $^7$Li/H$= (1.23^{+0.68}_{-0.32})\times 10^{-10}$~\citep{rya00} and $^6$Li/H$=(7.1\pm 0.7)\times 10^{-12}$~\citep{asp06}.  Thin solid lines around the lines of $d$($^6$Li) = 1 curve enclose the 1~$\sigma$ uncertainty in the adopted observational constraint based upon the dispersion of the observed plateau.  In the gray region, the condition $d$($^6$Li)$ > d$($^7$Li) is satisfied.\label{fig5}}
\end{figure}

In Fig.\ \ref{fig5}, the contours of $d$($^6$Li) [solid curves] and
$d$($^7$Li) [dashed curves] are shown.  The upper and lower solid curves
correspond to the abundance level which satisfies our adopted constraint
($1.7\times 10^{-12} \le ^6$Li/H $\le 7.1\times 10^{-11}$) as discussed
in subsec.~\ref{sec32} from the
abundance of MPHSs, ($^6$Li/H)$_{\rm MPHS}=(7.1\pm0.7)\times
10^{-12}$~\citep{asp06}.  Thus the upper right region of the figure is excluded for $^6$Li overproduction.
The right-upper side from the lower solid curve indicates the region for which the $^6$Li
abundance is higher than that observed in MPHSs.  

The three dashed curves correspond to $d$($^7$Li)=1.55, 2, 3, from right to
left, respectively.  In Fig.\ \ref{fig5} there is no $^7$Li over-destruction region where
    $^7$Li abundance is below the observed mean value.  $d$($^7$Li)=1.55 corresponds to the 1 sigma upper
    limit for MPHSs value~\citep{rya00}.  In the right-upper side
    from the 
dashed curve for $d$($^7$Li)=2, the resulting $^7$Li abundance is lower than $^7$Li/H$\approx 2.5 \times 10^{-10}$.  Therefore, we conclude that 
BBN with negatively charged particles provides a simultaneous solution to the $^7$Li
overproduction problem and the $^6$Li underproduction problem, [as also
deduced by~\citet{bir07}] in the parameter region
\begin{equation}
Y_X \gtrsim 0.9,~~~~~\tau_X\approx (1.0-1.8)\times 10^3~{\rm s}.
\label{eq20}
\end{equation}

For $\tau_X \gtrsim 10^5$~s and $Y_X \gtrsim 3$, the calculated abundance of $^7$Li increases slightly.  In this region the $^6$Li$_X(p,\gamma)^7$Be$_X$ reaction produces some amount of $^7$Be$_X$.
However, this parameter region is uninteresting due to an extreme overproduction of  $^6$Li  compared with observed abundances.  

In Fig.\ \ref{fig5}, there is a solid line corresponding to
$d$($^6$Li)=4.  The condition \\($^7$Li/H)/($^7$Li/H)$_{\rm MPHS} \lesssim$
($^6$Li/H)/($^6$Li/H)$_{\rm MPHS}$ or equivalently $d$($^7$Li)$\lesssim
d$($^6$Li) is satisfied in the gray region.  The gray colored region in
Fig.\ \ref{fig5} [cf.~Eq.\ (\ref{eq18})] is the most interesting and
relevant parameter region in order to solve both the $^6$Li and $^7$Li
problems.  The wider vertical band of $4 \times 10^{-6} \lesssim Y_X
\lesssim 8 \times 10^{-6}$ and $ \tau_X \gtrsim 10^4~{\rm s} $
corresponds to the solid box bounded by the vertical solid line in Fig.~\ref{fig4} and Eq.\ (\ref{eq17}). 
This region of the parameter space  solves only the $^6$Li problem,
but leaves the $^7$Li problem unresolved.

\begin{figure}
\plotone{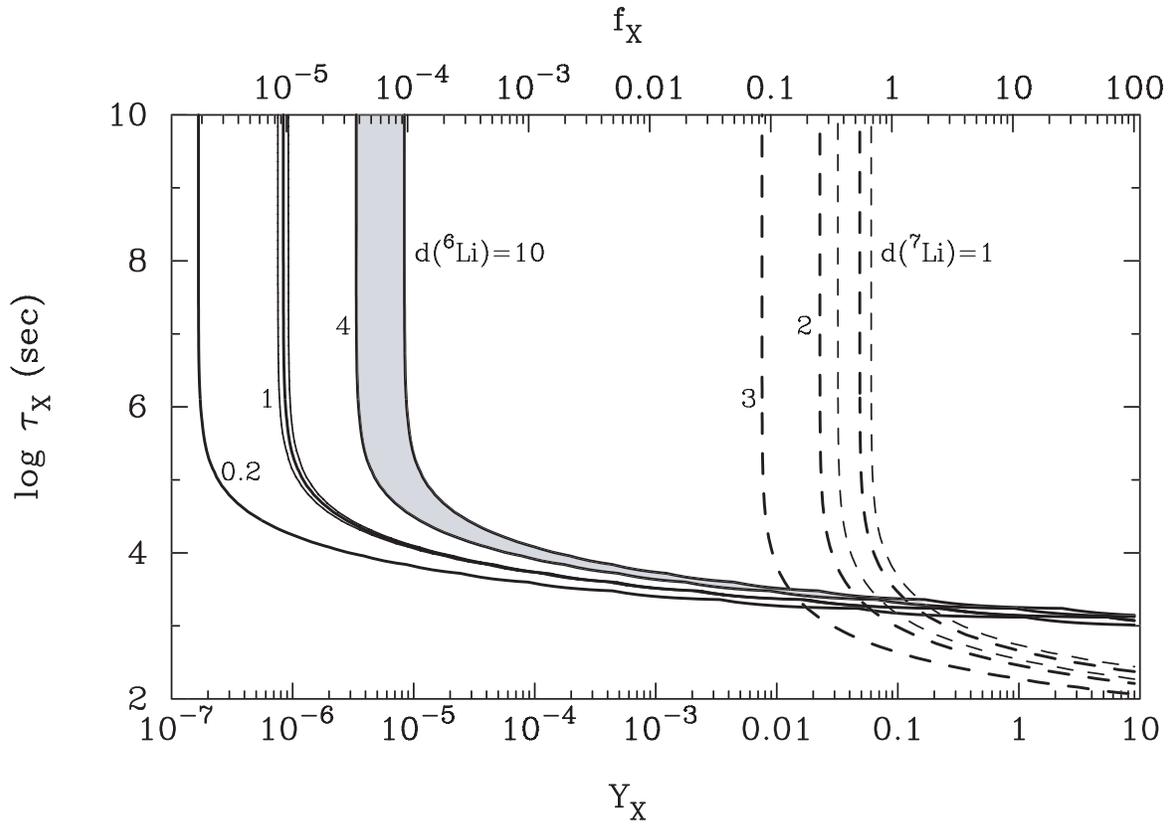}
\caption{Same as in Fig.\ \ref{fig5} when the charged-current decay of
 $^7$Be$_X \rightarrow ^7$Li$+X^0$ is included.  Thin dashed lines around the lines of $d$($^7$Li) = 1 curve enclose the 1~$\sigma$ uncertainty in the adopted observational constraint based upon the dispersion of the observed plateau.\label{fig6}}
\end{figure}

Finally, we consider another case where $^7$Be$_X$ converts to $^7$Li  by a weak
charged current transition from $X^-$ to $X^0$, i.e.  $^7$Be$_X \rightarrow ^7$Li+$X^0$.
This decay quickly transforms $^7$Be$_X$ to
$^7$Li~\citep[type II model in][]{bir07}.  In this model, the rate of
recombination effectively determines the $^7$Be$\rightarrow ^7$Li
conversion rate induced by $X^-$.  The results of this case are  shown in
Fig.\ \ref{fig6}.  The general features of Fig.\ \ref{fig6} are  very
similar to Fig.\ \ref{fig5}.  However, due to a slightly stronger destruction
rate due to  the  $^7$Be$_X \rightarrow ^7$Li+$X^0$ decay followed by the
$^7$Li($p$,$\alpha$)$^4$He and
$^7$Li($X^-$,$\gamma$)$^7$Li$_X$($p$,$\alpha$)$^4$He$_X$ reactions, the contours of the $^7$Li abundance are
systematically shifted toward smaller values of $Y_X$.  The parameter region which solves both the
$^6$Li and $^7$Li problems also slightly shifts to
\begin{equation}
Y_X \approx 0.04-0.2,~~~~~\tau_X \approx (1.4-2.6) \times 10^3~{\rm s}.
\label{eq21}
\end{equation}

\section{DISCUSSION}\label{sec4}

\subsection{Constraint on Dark Matter Particles}

If the dark matter particles, which we denote $Y^0$ in this article, are
the decay products of $X^-$ particles, the cosmological parameter
$\Omega_{\rm CDM}$ inferred from the WMAP-CMB data can be used to constrain
the mass of $Y^0$ when combined with our abundance constraints on the  $X^-$
particles from Eqs.\ (\ref{eq20}) and\ (\ref{eq21}).  We
suppose that $X^-$ decays to a dark-matter $Y^0$ particle and any residues, and
$Y_Y =Y_X$.  The WMAP-CMB constraint on $\Omega_{\rm CDM}=0.2$
corresponds to $Y_Y m_Y \lesssim 4.5$~GeV, i.e.
\begin{equation}
Y_X\approxeq Y_Y \lesssim \frac{4.5~{\rm GeV}}{m_Y},
\label{eq22}
\end{equation}
The calculated abundance constraints on $Y_X$ (Eqs. \ref{eq20} and \ref{eq22} can therefore be used to constrain the mass $m_Y$.  Note, that the calculated result in the present study does not particularly depend on the
assumed mass of the $X^-$.  In fact, only the second recombination rates and
the nuclear reaction rates between $X$-nuclei depend on the mass of
the $X^-$ (see discussion in subsec.~\ref{sec25}).  However, the main production and
destruction processes of $^6$Li, $^7$Li, and $^7$Be are completely free
from these processes.  We can thus consider the general constraint on
the number fraction of $X^-$ particles, $Y_X$.

The most interesting solution to both  the $^6$Li
underproduction and $^7$Li overproduction  involves the parameter space defined in
Eqs.~(\ref{eq20}) and~(\ref{eq21}).  Using  [Eq.~(\ref{eq20})] in which  one includes
the destruction reaction process $^7$Be$_X$+$p \rightarrow ^8$B$_X^{\ast {\rm a}}
\rightarrow ^8$B$_X$+$\gamma$~\citep{bir07}, the mass of the dark matter particle
$Y^0$ would be  constrained to be
\begin{equation}
m_Y \lesssim 5~{\rm GeV}~~.
\label{eq23}
\end{equation}
However, using  [Eq.~(\ref{eq21})]  in which  one includes
the $^7$Be$_X \rightarrow ^7$Li+$X^0$ process~\citep{bir07} the allowed
mass range increases to 
\begin{equation}
m_Y \lesssim 20-110~{\rm GeV},
\label{eq24}
\end{equation}

The long lifetime for $X^-$ decay is consistent with the mass of the decaying particle $X^-$ being close to the mass of the daughter
particle $Y^0$.  In this case one can deduce a constraint on the  $X^-$ mass of
\begin{equation}
m_X \lesssim O(100~{\rm GeV}).
\label{eq25}
\end{equation}

\subsection{Initial $X^-$ Abundance at BBN Epoch}

Here, we consider a simple estimation of the initial $X^-$ abundance after the
BBN epoch taking  $X^+$ and $X^-$ pair annihilation into account.  When
$X^+$ and $X^-$ particles are abundant at high temperature, both pair
creation and annihilation equilibrate.  However,  as the universe expands and
cools, the annihilation process proceeds until these particles freezeout at some  relic $X^-$
($X^+$) abundance.  A calculation of this is analogous to the well known weakly-interacting massive particle (WIMP) calculation of~\citet{kol90}.

Following their derivation, the  time evolution  the  $X$-to-photon ratio,  $\eta_X
\equiv n_X/n_\gamma=\eta Y_X$,  can be written,
\begin{equation}
\frac{d\eta_X}{dt}=-n_\gamma \langle \sigma v \rangle \eta_X^2.
\label{eq27}
\end{equation}
However, unlike the~\citet{kol90} calculation, the $X^-$ annihilation cross section is given 
by the electromaganetic formation of a positronium-like bound state  $(X^-,X^+)$
system, not by the  weak interaction associated  WIMPs.  Therefore,  the annihilation  cross section  does not scale the same as the WIMP annihilation rate, $\langle \sigma v \rangle_{WIMP} \propto (G_F^2 m_X^2)$, where $G_F$  is the Fermi coupling constant.  Rather, 
the much larger annihilation cross section~\citep{bir07} of interest here  is given by,
\begin{equation}
\langle \sigma v \rangle = \frac{2^{10} \pi^{3/2} \alpha^3}{3 \exp(4) m_X^{3/2} T^{1/2}}~~.
\label{eq28}
\end{equation}

The solution of Eq.~(\ref{eq27}) is then
\begin{equation}
\eta_X(T)=\left(\frac{1}{\eta_{X {\rm i}}} + 2H_{\rm i}^{-1} n_{\gamma {\rm i}} \langle \sigma v \rangle_{\rm i} \left(1-\left(\frac{T}{T_{\rm i}}\right)^{1/2}\right)\right)^{-1}.
\label{eq29}
\end{equation}
where the subscript i refers to values at some initial temperature $T_{\rm i}$.

Assuming that $\eta_{X{\rm i}} \sim O(1)$ and that the baryon-to-photon
ratio is $\eta=6.0 \times 10^{-10}$~\citep{spe07}, the freeze-out abundance of
$X^-$ can be written;
\begin{eqnarray}
Y_X &\approx & \frac{1}{\eta} \frac{1}{2H_{\rm i}^{-1} n_{\gamma {\rm i}} \langle \sigma v \rangle_{\rm i}} \nonumber \\
 &=& 0.0057 \left(\frac{g_\ast}{10.75}\right)\left(\frac{\eta}{6.0 \times 10^{-10}}\right)^{-1} \nonumber \\
&&\times \left(\frac{m_X}{50~{\rm GeV}}\right)^{3/2}\left(\frac{T_{\rm i}}{50~{\rm GeV}}\right)^{-1/2},
\label{eq30}
\end{eqnarray}
where $g_\ast$ is the total number of degrees of freedom of the relativistic
particles.  This approximation indicates that the charged massive
particles that existed in the early universe might remain as relics in
abundance of order $Y_X \sim 0.01$.  This is not very different from the
$Y_X$-value of the most interesting solution to the $^6$Li and $^7$Li
problems in Eq.\ (\ref{eq20}) or Eq.\ (\ref{eq21}) which we found, i.e.~the solution which leads to
the simultaneous destruction of $^7$Be and production of $^6$Li at levels that would produce concordance with observations.

\subsection{Direct Destruction of Nuclei by $X^-$ Decay}

Having shown how primordial $X^-$ particles could result in modified Li
isotope production, we need to consider what would happen to those
nuclei which retained an embedded $X^-$ particle when the $X^-$ particle
decays.  This decay was assumed by~\citet{kap06} to
interact sufficiently strongly with the host nucleus that it would induce nucleon emission.
For example, the decay could knock out a proton or neutron from $^4$He$_X$, producing
either $^3$He or $^3$H.  These nuclides could then interact with other
$^4$He nuclei to produce $^6$Li.  

However, this interaction was
found by those authors to occur too infrequently, which is consistent
with earlier studies~\citep[e.g.][]{ros75}.  It was therefore concluded that the resulting decay products would have little
effect if they were comprised only of leptons and photons.  Even so, they might
destroy some fraction of the nuclei in which the decays occurred if one of the decay products was a
pion~\citep{ros75,koh07}.  However, for the $X^-$ decays, the decay products might be expected to be
electroweakly interacting particles, and the $X^-$ decays might not be expected 
to significantly affect 
the host nuclei.  We therefore neglect decay-induced destruction in the present study.

\subsection{$^6$Li Production by Electromagnetic Energy Injection from
  $X^-$ Decay}

If the $X^-$ particles decay to charged leptons and any other residues, the
high-energy leptons thus produced could interact with background photons to lose
their energy and induce additional nonthermal nucleosynthesis.  In this section we argue that
such nucleosynthesis  (if it does occur) does not significantly alter  the parameter constraints deduced above.

There is also
a viable possibility that the decaying product is an electron.  Such a  high
energy electron quickly interacts with background photons ($e^\pm+\gamma
\rightarrow e^\pm+\gamma$ i.e. inverse Compton scattering).  The newly
produced energetic photons will also interact with background photons
($\gamma+\gamma \rightarrow e^- + e^+$ i.e. pair production).  In this
way high energy electrons can lose energy by making an electromagnetic cascade
shower.  Such a shower could trigger  non-thermal nucleosynthesis~\citep{dim88a,dim88b,dim89,jed00,cyb03,jed04,kaw05,kus06}.  

In~\citet{kus06} the  conditions on the
parameters of $X^-$ particles which lead  to a resolution of the $^6$Li underproduction problem
were delineated.  These are $Y_X m_X \geq Y_X E_{\rm EM} \sim
10^{-4}-10^{-3}$~GeV and $\tau_X \sim 10^8-10^{12}$~s, where $E_{\rm
EM}$ is the generated energy in the electromagnetic decay process.  The
constraint on $Y_X$ for a given $X^-$ mass $m_X$ is then
\begin{equation}
\frac{\left(10^{-4}-10^{-3}~{\rm GeV}\right)}{m_X} \lesssim Y_X.
\label{eq31}
\end{equation}

In the present paper we have considered  several destruction processes of 
$^7$Be$_X$.  Here we note that both Fig.\
\ref{fig5} and Fig.\ \ref{fig6} indicate the same constraint $Y_X
\lesssim 10^{-5}$ from $X^-$
BBN for $\tau_X \sim 10^8-10^{12}$~s which is the interesting lifetime
range in the
previous work based upon electromagnetic $X^-$ decay~\citep{kus06}.  The electromagnetic decay of
$X^-$ particles triggers late time $^6$Li production in amounts of $1
\lesssim d$($^6$Li)$\lesssim 10$ for these parameter regions.  Therefore, the compatible condition which avoids 
 an overproduction at the earlier epoch and still allows for non-thermal
production of $^6$Li at the later time is only $d$($^6$Li)$\lesssim 10$.
From this we deduce,
\begin{equation}
\frac{\left(10^{-4}-10^{-3}~{\rm GeV}\right)}{m_X} \lesssim 1 \times 10^{-5},
\label{eq32}
\end{equation}
which places a lower limit on the mass of the $X^-$ particle of
\begin{equation}
m_X \gtrsim 10-100~{\rm GeV}.
\label{eq33}
\end{equation}
When $m_X$ satisfies this lower limit, the electromagnetic decay of
$X^-$ particles can also operate at times long after the BBN epoch.
Clearly, electromagnetic and/or hadronic decays  may also contribute to
the final computed abundances.  And, in a subsequent work we propose to
investigate this.  However, for now our goal has been to clarify the
role of $X^-$ reactions and confirm that this effect alone can explain
the observed lithium abundances.  Note that in the case where the stable
daughter dark matter particle has nearly the same mass as the $X^-$
particle, there is not enough $Q$-value to produce a hadronic or
electromagnetic shower.

\section{CONCLUSIONS}\label{sec5}

We have investigated light-element nucleosynthesis during the big bang
in the presence  of massive, negatively-charged  $X^-$
particles.  Such particles would bind to light nuclei in the early universe.
As suggested by many authors, they would facilitate BBN by
enhancing the nuclear reaction rates  both by
reducing the charge of the bound $X$-nuclei, and by enabling
transfer reactions involving the $X^-$ particles.  We considered the
recombination processes of $X^-$ particles and normal nuclei.
Our conclusions are as follows:

First, as suggested in previous studies, the $X^-$ particles greatly enhance
the production of $^6$Li.  The main production process of $^6$Li is the
sequence of $^4$He$_X$ production through $X^-$ capture on $^4$He followed by  the $X^-$
transfer reaction $^4$He$_X$($d$,$X^-$)$^6$Li.  The resultant $^7$Li
abundance, however is almost the same as normal BBN value unless there
is a large $X^-$ abundance $Y_X \gtrsim 0.1$.

Secondly, when the lifetime of the $X^-$ particle is much longer than the
period of normal BBN, the $^6$Li abundance monotonically increases with
the $X^-$ particle abundance, except for very small $X^-$
abundance ($Y_X \lesssim 10^{-10}$).  On the other hand, the $^7$Li
abundance is nearly independent of  the $X^-$ particle abundance unless
 the $X^-$ particle abundance is larger than $\sim 0.1$ times the total
abundance of baryons.  In this case the $^7$Li abundance decreases with the $X^-$
particle abundance due to the resonance reaction of
$^7$Be$_X$($p$,$\gamma$)$^8$B$_X$ which reduces the $^7$Li abundance.

Thirdly, the $^6$Li/H and $^7$Li/H observed in MPHSs can
constrain the lifetime and abundance of an $X^-$ particle.  These
observational constraints  require the lifetime and abundance to be in the
ranges of $\tau_X \approx (1.0-1.8) \times 10^3$~s and $Y_X \gtrsim 0.9$.  When the reaction $^7$Be$_X \rightarrow ^7$Li+$X^0$ is also taken
into account, these ranges change to $\tau_X \approx (1.4-2.6)
\times 10^3$~s and $Y_X \approx 0.04-0.2$.  Therefore, introducing $X^-$
particles with an adequate lifetime and abundance can be a solution for
both of the factor of $\sim 1000$ underproduction of $^6$Li and the
factor of 3-4 overproduction of $^7$Li in standard BBN.

Fourthly, a constraint on the $X^-$ particle mass can be made from
the dark-matter content deduced from the WMAP analysis of the CMB.
If the abundance of $X^-$ particles is $Y_X\gtrsim 0.1-1$ as summarized
above, the mass of dark matter particles
$Y^0$, which are produced from the decay of $X^-$ particles, turns out to
be $m_Y\lesssim
10-100$~GeV, thus leading to the constraint $m_X\lesssim O(100~{\rm GeV})$
when $m_X \sim m_Y$.

In summary, although
several possible solutions have been proposed to solve the underproduction
problem of $^6$Li, they do not necessarily resolve the overproduction
problem of $^7$Li simultaneously.  \citet{pos07} and~\citet{cyb06} proposed the $X^-$ transfer reactions to produce $^6$Li and
$^7$Li.  Regarding $^6$Li production, however,~\citet{ham07}
have shown that the assumed reaction cross section
is not as large as the value deduced by Pospelov.  \citet{bir07} proposed
new destruction processes of $^7$Be$_X$ through atomic excitations, and
we here investigated yet another destruction process through a
$^8$B$^\ast_X$ nuclear excited
resonance in order to resolve the $^7$Li overproduction
problem.  The destruction efficiency of these newly proposed processes
depends on the excitation energies above the $p$+$^7$Be$_X$ separation
threshold.  We then found that a better estimation of the binding
energies leads to negligible effect from the nuclear excited resonance.
It would be useful, however, in future work to predict more precisely the
binding energies and excited states of exotic $X$-nuclei and their
reaction cross sections utilizing  a more realistic quantum mechanical
treatment.

\acknowledgments
We are very grateful to Professor Masayasu Kamimura for enlightening
suggestions on the nuclear reaction rates for transfer and radiative
capture reactions.  This work has been supported in part by the
Mitsubishi Foundation, the
Grant-in-Aid for Scientific Research (17540275) of the Ministry of
Education, Science, Sports and Culture of Japan, and the JSPS
Core-to-Core Program, International Research Network for Exotic Femto
Systems (EFES).  MK acknowledges the
support by the Japan Society for the Promotion of Science.  Work at the
University of Notre Dame was supported by the U.S. Department of Energy
under Nuclear Theory Grant DE-FG02-95-ER40934. RNB gratefully
acknowledges the support of the National Astronomical Observatory of Japan
during his stay there.

\clearpage

\begin{deluxetable}{cccc}
\tablecaption{Binding Energies of $X^-$ Particles to Nuclei\label{tab1}}
\tablewidth{0pt}
\tablehead{
\colhead{nuclide} & \colhead{$r_{\rm c}^{\rm RMS}$~(fm)\tablenotemark{a}} &
 \colhead{References} & \colhead{$E_{\rm Bind}$~(MeV)}}
\startdata
$^1$H$_X$  & 0.875 $\pm$ 0.007 & 1 & 0.025\\
$^2$H$_X$  & 2.116 $\pm$ 0.006 & 2 & 0.049\\
$^3$H$_X$  & 1.755 $\pm$ 0.086 & 3 & 0.072\\
$^3$He$_X$ & 1.959 $\pm$ 0.030 & 3 & 0.268\\
$^4$He$_X$ & 1.80  $\pm$ 0.04  & 4 & 0.343\\
$^6$Li$_X$ & 2.48  $\pm$ 0.03  & 4 & 0.806\\
$^7$Li$_X$ & 2.43  $\pm$ 0.02  & 4 & 0.882\\
$^8$Li$_X$ & 2.42  $\pm$ 0.02  & 4 & 0.945\\
$^6$Be$_X$ & 2.52  $\pm$ 0.02\tablenotemark{b}  & 4 & 1.234\\
$^7$Be$_X$ & 2.52  $\pm$ 0.02  & 4 & 1.324\\
$^8$Be$_X$ & 2.52  $\pm$ 0.02\tablenotemark{b}  & 4 & 1.401\\
$^9$Be$_X$ & 2.50  $\pm$ 0.01  & 4 & 1.477\\
$^7$B$_X$  & 2.68  $\pm$ 0.12\tablenotemark{c}  & 5 & 1.752\\
$^8$B$_X$  & 2.68  $\pm$ 0.12  & 5 & 1.840\\
$^9$B$_X$  & 2.68  $\pm$ 0.12\tablenotemark{c}  & 5 & 1.917\\
\enddata
\tablenotetext{a}{Root mean square charge radius}
\tablenotetext{b}{Taken from $^7$Be radius}
\tablenotetext{c}{Taken from $^8$B radius}
\tablecomments{References: 1= \citet{yao06}; 2= \citet{sim81};
 3=TUNL Nuclear Data, http://www.tunl.duke.edu/NuclData;
 4= \citet{tan88}; 5= \citet{fuk99}.}

\end{deluxetable}

\clearpage

\begin{deluxetable}{cccc}
\tablecaption{Nuclear Reaction Rates of $X$-Nuclides\label{tab2}}
\tablewidth{0pt}
\tablehead{
\colhead{Reaction} & \colhead{Reaction Rate (cm$^3$~s$^{-1}$~mole$^{-1}$)} &
 \colhead{Reverse Coefficient\tablenotemark{a}} & \colhead{Q~(MeV)}}
\startdata
$^3$He$_X$($n$,$\gamma$)$^4$He$_X$  & $7.2$ & 3.95 & 20.653\\
$^3$He$_X$($d$,$p$)$^4$He$_X$  & $3.9 \times 10^{10}T_9^{-2/3}\exp(-5.36/T_9^{1/3})$ & 8.49 & 18.428\\
$^6$Li$_X$($n$,$\gamma$)$^7$Li$_X$  & $5.2 \times 10^{3}$ & 1.48 & 7.327\\
$^6$Li$_X$($n$,$t$)$^4$He$_X$  & $1.7 \times 10^{8}$ & 0.577 & 4.321\\
$^6$Li$_X$($p$,$\gamma$)$^7$Be$_X$  & $5.5 \times 10^{5}T_9^{-2/3}\exp(-6.74/T_9^{1/3})$ & 1.48 & 6.124\\
$^6$Li$_X$($p$,$^3$He)$^4$He$_X$  & $2.7 \times 10^{10}T_9^{-2/3}\exp(-6.74/T_9^{1/3})$ & 0.577 & 3.557\\
$^7$Li$_X$($n$,$\gamma$)$^8$Li$_X$  & $5.2 \times 10^{3}$ & 1.58 & 2.096\\
$^7$Li$_X$($p$,$\gamma$)$^8$Be$_X$  & $1.3 \times 10^5T_9^{-2/3}\exp(-6.74/T_9^{1/3})$ & 7.89 & 17.774\\
$^7$Li$_X$($p$,$\alpha$)$^4$He$_X$  & $9.0 \times 10^8T_9^{-2/3}\exp(-6.74/T_9^{1/3})$ & 1.00 & 16.808\\
$^8$Li$_X$($p$,$\gamma$)$^9$Be$_X$  & $1.0 \times 10^5T_9^{-2/3}\exp(-4.25/T_9^{1/3})$ & 2.47 & 17.420\\
$^6$Be$_X$($n$,$\gamma$)$^7$Be$_X$  & $5.2 \times 10^{3}$ & 0.493 & 10.766\\
$^6$Be$_X$($n$,$p$)$^6$Li$_X$  & $2.7 \times 10^{9}$ & 0.333 & 4.642\\
$^7$Be$_X$($n$,$\gamma$)$^8$Be$_X$  & $5.2 \times 10^{3}$ & 7.89 & 18.977\\
$^7$Be$_X$($n$,$p$)$^7$Li$_X$  & $2.7 \times 10^{9}$ & 1.00 & 1.202\\
$^7$Be$_X$($p$,$\gamma$)$^8$B$_X$  & $1.6 \times 10^8 T_9^{-2/3}\exp(-8.86/T_9^{1/3})$ & 1.58 & 0.654\\
&$+1.6 \times 10^6 T_9^{-3/2}\exp(-1.92/T_9)$ &&\\
$^7$Be$_X$($d$,$p$)$^8$Be$_X$  & $8.9 \times 10^{11} T_9^{-2/3}\exp(-11.13/T_9^{1/3})$ & 16.97 & 16.752\\
$^7$Be$_X$($d$,$p$$\alpha$)$^4$He$_X$  & $8.9 \times 10^{11}T_9^{-2/3}\exp(-11.13/T_9^{1/3})$ & 2.15 & 15.786\\
$^8$Be$_X$($n$,$\gamma$)$^9$Be$_X$  & $5.2 \times 10^{3}$ & 2.47 & 1.741\\
$^8$Be$_X$($p$,$\gamma$)$^9$B$_X$  & $1.4 \times 10^6 T_9^{-3/2}\exp(-5.60/T_9)$ & 2.47 & 0.331\\
$^9$Be$_X$($n$,$\gamma$)$^{10}$Be$_X$  & $5.2 \times 10^{3}$ & 7.89 & 6.913\\
$^9$Be$_X$($p$,$\gamma$)$^{10}$B$_X$  & $1.2 \times 10^7T_9^{-2/3}\exp(-8.83/T_9^{1/3})$ & 1.13 & 7.145\\
$^{10}$Be$_X$($p$,$\gamma$)$^{11}$B$_X$  & $1.2 \times 10^7T_9^{-2/3}\exp(-8.83/T_9^{1/3})$ & 0.493 & 11.750\\
$^9$B$_X$($n$,$\gamma$)$^{10}$B$_X$  & $5.2 \times 10^{3}$ & 1.13 & 8.556\\
$^9$B$_X$($p$,$\gamma$)$^{10}$C$_X$  & $1.9 \times 10^5T_9^{-2/3}\exp(-10.70/T_9^{1/3})$ & 7.89 & 4.727\\
$^{10}$B$_X$($n$,$\gamma$)$^{11}$B$_X$  & $5.2 \times 10^{3}$ & 3.45 & 11.517\\
$^{10}$B$_X$($p$,$\gamma$)$^{11}$C$_X$  & $1.9 \times 10^5T_9^{-2/3}\exp(-10.70/T_9^{1/3})$ & 3.45 & 9.366\\
$^{11}$B$_X$($p$,$\gamma$)$^{12}$C$_X$  & $1.9 \times 10^7 T_9^{-2/3}\exp(-10.70/T_9^{1/3})$ & 7.89 & 16.638\\
$^{10}$C$_X$($n$,$\gamma$)$^{11}$C$_X$  & $5.2 \times 10^{3}$ & 0.493 & 13.195\\
$^{11}$C$_X$($n$,$\gamma$)$^{12}$C$_X$  & $5.2 \times 10^{3}$ & 7.89 & 18.789\\
$^4$He$_X$($d$,$\gamma$)$^6$Li$_X$  & $2.2 \times 10^1 T_9^{-2/3}\exp(-5.35/T_9^{1/3})$ & 2.79 & 1.936\\
$^3$He$_X$($\alpha$,$\gamma$)$^7$Be$_X$  & $2.9 \times 10^6 T_9^{-2/3}\exp(-10.70/T_9^{1/3})$ & 3.95 & 2.642\\
$^4$He$_X$($t$,$\gamma$)$^7$Li$_X$  & $2.0 \times 10^5 T_9^{-2/3}\exp(-6.12/T_9^{1/3})$ & 2.56 & 3.006\\
$^4$He$_X$($^3$He,$\gamma$)$^7$Be$_X$  & $3.2 \times 10^6 T_9^{-2/3}\exp(-9.72/T_9^{1/3})$ & 2.56 & 2.567\\
$^4$He$_X$($^6$Li,$\gamma$)$^{10}$B$_X$  & $2.4 \times 10^{6} T_9^{-2/3}\exp(-16.05/T_9^{1/3})$ & 6.22 & 6.154\\
$^9$Be$_X$($p$,$^6$Li)$^4$He$_X$  & $1.9 \times 10^{11} T_9^{-2/3}\exp(-8.83/T_9^{1/3})$ & 0.181 & 0.991\\
$^4$He$_X$($\alpha$,$\gamma$)$^8$Be$_X$  & $2.9 \times 10^6 T_9^{-2/3}\exp(-10.70/T_9^{1/3})$\tablenotemark{b} & 7.89 & 0.966\\
$^6$Li$_X$($\alpha$,$\gamma$)$^{10}$B$_X$  & $3.0 \times 10^{6} T_9^{-2/3}\exp(-16.98/T_9^{1/3})$ & 3.38 & 5.692\\
$^7$Li$_X$($\alpha$,$\gamma$)$^{11}$B$_X$  & $2.7 \times 10^{7} T_9^{-2/3}\exp(-16.98/T_9^{1/3})$ & 7.89 & 9.882\\
$^7$Be$_X$($\alpha$,$\gamma$)$^{11}$C$_X$  & $6.6 \times 10^{7} T_9^{-2/3}\exp(-22.26/T_9^{1/3})$ & 7.89 & 8.934\\
$^4$He$_X$($d$,$X^-$)$^6$Li  & $2.37 \times 10^8 (1 -0.34 T_9)T_9^{-2/3}\exp(-5.33/T_9^{1/3})$  & 0.192 & 1.131\\
$^6$Be$_X$(,$e^+$$\nu_e$)$^6$Li$_X$  & $T_{1/2}=2.3$~s &  & 3.349\\
$^8$B$_X$(,$e^+$$\nu_e$)$^8$Be$_X$  & $T_{1/2}=1.2$~s &  & 17.030\\

\enddata
\tablenotetext{a}{For nuclides $a=i, j, k, ... $ with mass numbers $A_a$
 and spins $g_a$, the reverse coefficients are defined as
 follows~\citep{fow67} on the assumption that $X^-$ particle is much
 heavier than nuclides: $0.9867(g_i g_j/g_k)A_j^{3/2}$ for a process
 $i_X$($j$,$\gamma$)$k_X$, $(g_i g_j/(g_k g_l))(A_j/A_k)^{3/2}$ for a
 process $i_X$($j$,$k$)$l_X$, and $1.0134 (g_i g_j/(g_k g_l g_m))(A_j/(A_k A_l))^{3/2}$ for a process $i_X$($j$,$k$$l$)$m_X$.}
\tablenotetext{b}{The $S$-factor for the reaction $^3$He($\alpha$,$\gamma$)$^7$Be was used}

\end{deluxetable}

\end{document}